\documentclass{aa} 

\usepackage{graphicx}
\usepackage{txfonts}
\usepackage{hyperref}
\usepackage{placeins}

\usepackage{natbib}
\bibpunct{(}{)}{;}{a}{}{,} 

\begin{document}

   \title{Non-radial oscillations mimicking a brown dwarf orbiting the cluster giant NGC 4349 No. 127}


   \author{Dane Spaeth\thanks{Fellow of the International Max Planck Research School for Astronomy and Cosmic Physics at the University of Heidelberg (IMPRS-HD)}
          \and
          Sabine Reffert
          \and
          Emily L. Hunt
          \and
          Adrian Kaminski
          \and
          Andreas Quirrenbach}

   \institute{ Landessternwarte, Zentrum für Astronomie der Universität Heidelberg, Königstuhl 12, 69117    Heidelberg, Germany \\
   \email{dane.spaeth@lsw.uni-heidelberg.de}
             }

   \date{Received 28 March 2024 / 
   Accepted 09 July 2024}

 
  \abstract
   {Several evolved stars have been found to exhibit long-period radial velocity variations that cannot be explained by planetary or brown dwarf companions. Non-radial oscillations caused by oscillatory convective modes have been put forth as an alternative explanation, but no modeling attempt has yet been undertaken.}
   {We provide a model of a non-radial oscillation, aiming to explain the observed variations of the cluster giant NGC 4349 No. 127. The star was previously reported to host a brown dwarf companion, but whose existence was later refuted in the literature.}
   {We reanalyzed 58 archival HARPS spectra of the intermediate-mass giant NGC 4349 No. 127. We reduced the spectra using the SERVAL and RACCOON pipelines, acquiring additional activity indicators. We searched for periodicity in the indicators and correlations between the indicators and radial velocities. We further present a simulation code able to produce synthetic HARPS spectra, incorporating the effect of non-radial oscillations, and compare the simulated results to the observed variations. We discuss the possibility that non-radial oscillations cause the observed variations. }
   { We find a positive correlation between chromatic index and radial velocity, along with closed-loop Lissajous-like correlations between radial velocity and each of the spectral line shape indicators (full width at half maximum, and contrast of the cross-correlation function and differential line width). Simulations of a low-amplitude, retrograde, dipole ($l=1, m=1$), non-radial oscillation can reproduce the observed behavior and explain the observables. Photometric variations below the detection threshold of the available ASAS-3 photometry are predicted. The oscillation and stellar parameters are largely in agreement with the prediction of oscillatory convective modes.}
   {The periodic variations of the radial velocities and activity indicators, along with the respective phase shifts, measured for the intermediate-mass cluster giant NGC 4349 No. 127, can be explained by a non-radial oscillation.}

   \keywords{ stars: individual: NGC 4349 No. 127 -- stars: oscillations -- stars: evolution -- planets and satellites: detection -- techniques: radial velocities}

   \maketitle

\section{Introduction}
With the advent of large-scale transit surveys such as {\it Kepler} \citep{Borucki2010}, K2 \citep{Howell2014}, and TESS \citep{Ricker2015}, the number of known exoplanets has increased drastically to over 5600 confirmed planets to date.\footnote{\url{https://exoplanetarchive.ipac.caltech.edu/} (June 2024)} Despite this immense wealth of newly discovered planets, less than 200 of these have been found orbiting evolved stars.\footnote{\url{https://www.lsw.uni-heidelberg.de/users/sreffert/giantplanets/giantplanets.php}} Due to intrinsic biases of the transit surveys, the majority of planets orbiting giant stars have been found using the radial velocity (RV) method. 

Targeting giant stars in RV surveys offers important additions to surveys targeting main-sequence stars. Specifically, it allows the detection of planets around the evolved counterparts of stars more massive than about $1.5\,\mathrm{M_\sun}$  \citep{Reffert2015}, which are increasingly inaccessible to RV measurements during their main-sequence phase, due to high effective temperatures and rotation rates \citep{Sato2003, Galland2005, Johnson2007, Lagrange2009, Assef2009}. Furthermore, analyzing the planet population in later evolutionary stages allows  studies of the effect of stellar evolution on planetary systems. Therefore, detecting a statistically meaningful sample of planets around giant stars is crucial in order to capture a full picture of the general planet population. For simplicity, we also refer to low-mass brown dwarfs ($M \lesssim 30 M_\mathrm{Jup}$) as planets within this work.   

However, detecting planets orbiting giants using the RV method has its own challenges. First, most giant stars are known to undergo short-term, stochastically driven p-mode oscillations, {called} solar-like oscillations. For giants, these typically occur on timescales of hours to days and lead to RV jitter on the order of $\sim 10\,\mathrm{m\,s^{-1}}$ to $\sim 20\,\mathrm{m\,s^{-1}}$, even for giant stars considered to be relatively stable \citep{Kjeldsen1995, Hekker2006a, Hekker2008}. This intrinsic noise can often be overcome with sufficient statistics, but limits the detectable orbital companions to those with rather high masses (typically Jovian or higher). 

Less understood, and therefore more challenging, is the recent discovery of several giant stars showing long-term, periodic RV variations, some of which were already attributed to planets or brown dwarfs, yet were later shown to be incompatible with orbital companions. These include \object{$\gamma$ Dra} \citep{Hatzes2018}, \object{Aldebaran} \citep{Reichert2019}, \object{$\epsilon$ Cyg} \citep{Heeren2021}, \object{42 Dra} \citep{Dollinger2021}, Sanders 364\footnote{SIMBAD identifier: \object{BD+12 1917}} \citep{Zhou2023}, \object{41 Lyn}, \object{14 And} (both \citealt{Teng2023b}), and the cluster giants NGC 2423 No. 3, NGC 2345 No. 50, NGC 3532 No. 670, and NGC 4349 No. 127\footnote{SIMBAD identifiers: \object{NGC 2423 3}, \object{NGC 2345 50}, \object{HD 96789}, and \object{NGC 4349 127}} \citep{DelgadoMena2018, DelgadoMena2023}. \citet{Dollinger2021} show that doubtful planet detections around giant stars mostly accumulate at periods between 300~d and 800~d and around stars with radii $R > 21\,R_\sun$, pointing at a common origin of these RV variations.

\citet{Wolthoff2022} present the planet occurrence rate around evolved stars as a function of the orbital period, which follows a broken power law relation peaking at $P \sim 720\,\mathrm{d}$, very close to the periods of the identified false positive detections. If a significant number of as-yet-unidentified false positives contaminate the comparably small sample of planets around giants, significant implications on the planet occurrence rate and its interpretation are to be expected \citep{Wolthoff2022}.

Moreover, as the phenomenon is still poorly understood and can mimic planets quite convincingly in RV data, it is often impracticable to unambiguously confirm new planets around giant stars (see, e.g., \citealt{TalaPinto2020, Niedzielski2021, Jeong2022, Teng2022b, Zhou2023}). This is due to the fact that it is not yet clear which observables are the most decisive diagnostics to separate intrinsic signals from orbital companions for giants. This problem is made even worse by the varying availability, in data sets from different spectrographs, of spectral diagnostics targeting such intrinsic variations. These are commonly referred to as activity indicators, and we adopt this term even though we do not (mainly) target magnetic activity in this work. 

So far, several intrinsic origins have been discussed. Radial oscillations can usually be ruled out due to the long-period nature of the suspected signals, which appear at periods much longer than the fundamental radial mode \citep{Hatzes1993, Cox1972, Hatzes2018, Reichert2019}. In the binary system $\epsilon$ Cyg, the heartbeat phenomenon was considered \citep{Heeren2021}; however, it  is  not applicable to single stars. Magnetic surface structures, such as cool spots, can in most cases equally be ruled out, as they should manifest themselves in much larger photometric variations than were reported for the above stars \citep{Reichert2019, Heeren2021}. \citet{DelgadoMena2023} (and references therein) propose magnetic plages or other magnetic structures that locally reduce convection and could cause RV variations without associated photometric variations. However,  the magnetic fields on the surface of most giant stars are still  not well understood  (see, e.g., \citealt{Auriere2011, Auriere2013, Auriere2015, Konstantinova-Antova2024}, for notable exceptions). 

Non-radial oscillations, on the other hand, can cause large amplitude RV variations at long periods \citep{Hatzes1996, Hatzes1999}. \citet{Hatzes2018} link the aforementioned phenomenon of false positive planet detections to oscillatory convective modes presented by \citet{Saio2015}. These modes were proposed as the origin for the  sequence D of the long secondary periods (LSPs) observed for variable bright giant stars \citep{Wood1999, Wood2000}. These are dipole ($l=1$) $g^{-}$-modes that become oscillatory in the non-adiabatic conditions present in the envelopes of luminous ($\log L/L_\sun \gtrsim 3$) giant stars \citep{Takayama2020}. \citet{Reichert2019} build on this argument, showing in their Fig.~8 that both $\gamma$ Dra and Aldebaran fall into a region in the period-luminosity diagram in which an extrapolation of the models by \citet{Saio2015} could potentially explain the observed variations. 
Nevertheless, no satisfying attempt to apply models of non-radial oscillations to the observables of a known false positive evolved planet host has been published to date.

In this work we reexamine the RV variations of the cluster giant NGC 4349 No. 127, previously thought to host a brown dwarf companion in a $P=677.8\,\mathrm{d}$ orbit originally published by \citet{Lovis2007}. However, \citet{DelgadoMena2018} show that, while the RV signal is stable in HARPS (High Accuracy Radial velocity Planet Searcher) spectra, variations of the full width at half maximum (FWHM) of the cross-correlation function (CCF) and the $\mathrm{H\alpha}$ index at the orbital period are present. The authors therefore refute the companion's existence, favoring rotational modulation of magnetic activity as the most likely alternative, while not ruling out non-radial oscillations. \citet{DelgadoMena2023} affirm these findings by presenting 11 additional RV measurements.

Here, we present a reanalyis of the HARPS data set adding measurements of the chromatic index (CRX) and differential line width (dLW) variations by using the SERVAL \citep{Zechmeister2018} pipeline, as well as the contrast of the CCF by using the RACCOON \citep{Lafarga2020} reduction software. We reveal a linear correlation between the RV, on the one hand, and the CRX and the $\mathrm{H\alpha}$ indicator, on the other hand, respectively. We further show that the dLW, the FWHM of the CCF, and the contrast of the CCF each correlate with the RV in a ``closed-loop'' relation. We additionally present simulations of the observational effects of non-radial oscillations that can closely reproduce the periods and amplitudes of the activity indicator variations, as well as the phase differences with the RV.

This paper is structured as follows. In Sect.~\ref{sec:observations} we give an overview of the stellar parameters, observations, and data reduction. In Sect.~\ref{sec:simulations} we present the \texttt{pyoscillot} simulation suite developed to simulate the effect of non-radial oscillations on HARPS spectra.  We present newly detected correlations between the activity indicators and the RVs in Sect.~\ref{sec:results}, and show that they are consistent with a retrograde, dipole, non-radial oscillation model.
In Sect.~\ref{sec:othermodes} we present general observational properties of different oscillation modes. We discuss our non-radial oscillation model in the context of oscillatory convective modes in Sect.~\ref{sec:discussion}, before summarizing our findings in Sect.~\ref{sec:summary}.

\section{Observations and stellar parameters}
\label{sec:observations}
\subsection{Stellar parameters}
NGC 4349 No. 127 is the most evolved star of the open cluster \object{NGC 4349} \citep{DelgadoMena2018}, located at a distance $d = 1788.9\pm2.9\,\mathrm{pc}$ \citep{Hunt2023}, with an estimated age of around $300\,\mathrm{Myr}$ \citep{Holanda2022, Tsantaki2023, Hunt2023}. The stellar parameters vary considerably between different studies and are summarized in Table~\ref{tab:params}. 

\begin{figure}
    \centering
    \includegraphics{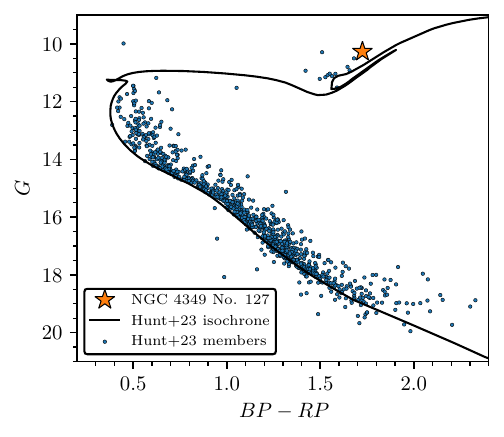}
    \caption{Color-magnitude diagram for the open cluster NGC 4349 based on {\it Gaia} DR3 photometry. The membership list and isochrone fit were taken from \citet{Hunt2023}. NGC 4349 No. 127 is the most evolved star of the cluster,  indicated by the orange star marker. Its evolutionary state is ambiguous between the first or second ascent on the RGB.}
    \label{fig:cmd}
    
\end{figure}

In Fig.~\ref{fig:cmd}, we plot a color-magnitude diagram based on {\it Gaia} DR3 photometry \citep{GaiaCollaboration2016, GaiaCollaboration2023}, highlighting the position of NGC 4349 No. 127 as the orange star marker. The membership list and isochrone fit were taken from \citet{Hunt2023}. The star's exact evolutionary state is ambiguous between the first or second ascent on the red giant branch (RGB), as was discussed in the literature \citep{DelgadoMena2016, Tsantaki2023}. One reason for this ambiguity is the differential extinction $0.57^{+0.30}_{-0.31}\,\mathrm{mag}$ in the V band present in the cluster \citep{Hunt2023}, which is also evident from the spread at the turn-off of the main sequence in Fig.~\ref{fig:cmd}. This complicates the determination of a specific extinction value for NGC 4349 No. 127, adding uncertainty to its intrinsic photometry. The difficulty in determining an accurate extinction value could also (partially) explain the variability of the stellar parameters summarized in Table~\ref{tab:params}.

We also tested the stellar parameters using Bayesian inference based on {\it Gaia} DR3 parallaxes and photometry using \texttt{SPOG+}\footnote{\url{https://github.com/StephanStock/SPOG}} \citep{Stock2018}. We use extinction values taken from the Starhorse catalog \citep{Anders2022}. The metallicity was taken from \citet{Tsantaki2023}. The tool yields a probability of $99.1\%$ for the star being on the horizontal branch (HB) or second ascent on the RGB. The stellar parameters derived by \texttt{SPOG+} are listed in Table~\ref{tab:params}.

While the results for $T_\mathrm{eff}$ are roughly consistent with other determinations in the literature, \texttt{SPOG+} estimates a significantly lower mass, radius, and luminosity. The age determination of the star $\tau = 930^{+300}_{-330}\,\mathrm{Myr}$ also significantly deviates from the estimated age for the whole cluster from other studies. We assume that stellar parameters derived from spectroscopy or ones that take the whole cluster into account should generally be regarded as more accurate. \texttt{SPOG+} is also quite sensitive to the choice of extinction values, which are difficult to determine. However, as masses of giant stars derived from evolutionary models are suspected to be overestimated (see, e.g., \citealt{Lloyd2011}), we note this somewhat lower mass estimate, as well as the lower mass estimates by \citet{Anders2022} and \citet{Mortier2013a}, and the general variability of the parameters in Table~\ref{tab:params}.

The star has furthermore been found to show enhanced Li abundance compared to other giant stars in the cluster and compared to field stars \citep{Carlberg2016, DelgadoMena2016, Tsantaki2023}, which was proposed to be caused by planet engulfment. However, \citet{Holanda2022} notes that the determined Li abundance is still below the traditional limit for Li-rich giant stars $\log \epsilon(\textrm{Li}) \geq 1.50$. 

Given the precision of the results and favoring parameters derived by spectroscopy, we base our analysis and discussion on the stellar parameters derived by \citet{Tsantaki2023} and \citet{DelgadoMena2023}. For the simulations presented in Sect.~\ref{sec:results}, we chose $(T_\mathrm{eff}=4500\,\mathrm{K}, \log g = 2.0, [\mathrm{Fe/H}]=0.0)$ as the closest grid point of the PHOENIX spectral library. We refrain from interpolating the base spectrum of the simulations to the exact parameters due to the differences shown in the literature (Table~\ref{tab:params}) and to avoid interpolation errors. Using the relation by \citet{Hekker2007}, we determine the value of the macroturbulence to be $\zeta=4822\pm39\,\mathrm{m\,s^{-1}}$ at this effective temperature.

\begin{table*}
\centering
\caption{Overview of the stellar parameters determined for NGC 4349 No. 127 in the literature.}
\label{tab:params}
\resizebox{\textwidth}{!}{
\begin{tabular}{lllllllllll}
\hline
$M$                     & $R$            & $L$              & $T_\mathrm{eff}$   & $\log g [\mathrm{cm\,s^{-2}}]$               & $[\mathrm{Fe/H}]$ & $\xi$                & $v \sin i$           & $\zeta$             & $A_V$ & Ref.              \\
$M_\sun$               & $R_\sun$     & $L_\sun$       & $\textrm{K}$       &   & dex               & $\mathrm{km\,s^{-1}}$ & $\mathrm{km\,s^{-1}}$ & $\mathrm{km\,s^{-1}}$ &  mag    &             \\ \hline
$2.20^{+0.38}_{-0.20}$ & $23.3^{+0.9}_{-1.4}$ & $211^{+3}_{-4}$ & $4566^{+123}_{-70}$ & $2.04^{+0.13}_{-0.06}$ & -                 & -                    & -                    & -            &    $0.67^{+0.06}_{-0.07}$     & \texttt{SPOG+}, 1 (phot) \\
$3.29^{+0.28}_{-0.56}$ &    -          &      -          & -                  & -                      & -                 & -                    & -                    & -                  & $0.86^{+0.19}_{-0.15}$ & 2 (phot)                \\
$3.01 \pm 0.24$        & - & -       & $4417 \pm 12$      & $1.78 \pm 0.05$        & $0.17 \pm 0.02$  & -                    & -                 & -             & 1.16   & 3 (spec)       \\
$3.41 \pm 0.27$        &     -         &     -           & $4567 \pm 31$      & $1.83 \pm 0.08$        & -                 & -                    & -                    & -                  & 1.16 & 3 (phot)          \\

-  & $38.0\pm2.6$ & $575.4$       & -      & -        & -  & -                    & 4.81                 & 4.66              &  1.16 & 4, 3 (spec+phot)       \\

-                      &        -      &          -      & $4420 \pm 70$      & $1.70 \pm 0.20$        & $-0.13 \pm 0.13$  & $1.51\pm0.07$                    & $4.20 \pm 0.60$      & $3.0$             & $1.24\pm0.06$  & 5 (spec)          \\
-                      &    -          &        -        & $4597 \pm 39$      & $1.78 \pm 0.09$        & -                 & -                   & -                    & -             & $1.24\pm0.06$      & 5 (phot)          \\
$1.56^{+0.20}_{-0.36}$ & - & - & $4199^{+61}_{-57}$ & $1.43^{+0.01}_{-0.03}$ & $-0.29^{+0.01}_{-0.11}$ & - & - & - & $0.67^{+0.06}_{-0.07}$ & 6 (phot)\\
$2.4 $            &   -           &     -           & $4370 \pm 78$      & $1.70 \pm 0.22$        & $-0.20 \pm 0.11$  & $1.70 \pm 0.06$      & $<2$                 & $\sim 5.0$          & 0.90 & 7 (spec)                \\
$3.81 \pm 0.23$        & $37.0\pm4.9$   & $646^{+205}_{-156}$         & $4503 \pm 70$      & $1.99 \pm 0.19$        & $-0.13 \pm 0.04$  & $1.77 \pm 0.07$      & $6.13$               & -   &            -     & 8   (spec+phot)              \\
$1.37 \pm 0.37$        &   -          &        -        & $4445 \pm 87$      & $1.64 \pm 0.23$        & $-0.25\pm0.06$    & $1.84\pm0.08$        & -                    & -                   & - & 9  (spec+phot)               \\
$3.77 \pm 0.36$        &   $44.7\pm2.5$           &   $774^{+467}_{-291}$             & $4519 \pm 100$     & $1.92 \pm 0.2$         & $-0.21 \pm 0.12$  & $2.08 \pm 0.20$      & -                    & -                 & $1.08$  & 10 (spec+phot)                \\ \hline
\end{tabular}}
\tablefoot{Presented are mass $M$, radius $R$, luminosity $L$, effective temperature $T_\mathrm{eff}$, surface gravity $\log g$, metallicity $[\mathrm{Fe/H}]$, microturbulent velocity $\xi$, rotational broadening $v \sin i$,  macroturbulent velocity $\zeta$, and the V-band extinction $A_V$ used to determine the parameters. The values for reddening $E(B-V)$ given by \citet{Holanda2022} and \citet{Carlberg2016} were converted assuming $A_V=3.1 E(B-V)$.}
\tablebib{(1)~\citet{Stock2018}; (2)~\citet{Hunt2024}; (3)~\citet{Tsantaki2023}; (4)~\citet{DelgadoMena2023}; (5)~\citet{Holanda2022}; (6)~\citet{Anders2022}; (7)~\citet{Carlberg2016}; (8)~\citet{DelgadoMena2016}; (9)~\citet{Mortier2013a}; (10)~\citet{Ghezzi2010} }
\end{table*}

\subsection{Data}

We downloaded 58 publicly available, extracted HARPS \citep{Mayor2003} spectra from the ESO science archive\footnote{\url{http://archive.eso.org/scienceportal/home}} for NGC 4349 No. 127. The spectra were taken between 2005 and 2022 and have signal-to-noise ratios (S/N) per extracted pixel between 25 and 62 (with one outlier at 14.7) at the peak of the blaze function of (physical) order 102, which is centered around $6000\,\textrm{\AA}$. The data were taken as part of a long-term RV survey of intermediate-mass giant stars in open clusters and were originally published by \citet{Lovis2007} and \citet{DelgadoMena2018, DelgadoMena2023}.

We reduced the spectra using the SERVAL pipeline \citep{Zechmeister2018}, which derives radial velocities via a least-squares optimization of the RV shift relative to a high signal-to-noise stellar template obtained by coadding the available observations. SERVAL was shown to produce slightly more precise RV results than the default HARPS data reduction software (DRS) \citep{Trifonov2020}. The SERVAL pipeline furthermore computes additional activity indicators, namely the CRX and the dLW \citep{Zechmeister2018}, which have proven to be effective indicators to detect activity for main-sequence dwarfs, especially in context of the CARMENES (Calar Alto high-Resolution search for M dwarfs with Exoearths with Near-infrared and optical Echelle Spectrographs) survey \citep{Quirrenbach2014}. It moreover calculates several line indices, of which the $\mathrm{H}\alpha$ index will be used within this work.  

We furthermore utilized the RACCOON reduction software \citep{Lafarga2020} to obtain radial velocities using a CCF with a weighted binary mask. The mask was created from the SERVAL template via the routines provided by the RACCOON software. It furthermore provides three activity indicators targeting the contrast and FWHM of the CCF and calculates the Bisector Inverse Slope (BIS) \citep{Lafarga2020}. We prefer the RACCOON pipeline over the default HARPS DRS, since the latter is not publicly available and can therefore not be used to reduce the simulated data presented in Sect.~\ref{sec:results}. The results of the RACCOON pipeline are generally in good agreement with the HARPS DRS results, but yield somewhat higher RV uncertainties compared to HARPS DRS and SERVAL (see Table~\ref{tab:data}). 

We separated the spectra taken before (46 spectra) and after (12 spectra) 2 June 2015,  to account for the HARPS fiber change.\footnote{\url{https://www.eso.org/sci/facilities/lasilla/instruments/harps/news.html}} We did not separate the spectra further, accounting for the HARPS warm-up on 23 March 2020, as only three spectra were taken after this date. We treated the two subsets as separate RV time series, allowing for a relative offset. 

We restricted the analysis of the activity indicators to the 46 spectra acquired before the HARPS fiber exchange, as we noted significant offsets of the activity indicators by analyzing the activity time series of quiet stars using the HARPS RVBANK \citep{Trifonov2020} (see also Appendix A of \citealt{DelgadoMena2023}). As these offsets are (to our knowledge) poorly understood and constrained, we decided to focus merely on the spectra acquired prior to the fiber change. Both the SERVAL and RACCOON RVs and activity indicators are listed in Table~\ref{tab:data}. 

\section{Simulations}
\label{sec:simulations}
To study whether non-radial oscillations could explain the variations of the radial velocities and the activity indicators, we simulate the observational effect of these on HARPS spectra. The simulations are part of the simulation suite \texttt{pyoscillot}.\footnote{\url{https://github.com/DaneSpaeth/pyoscillot}}

\subsection{Description of non-radial oscillations}
We start out with a model of the stellar photosphere, using spherical coordinates ($\theta$,~$\phi$), with $\theta$ being the polar angle (or colatitude), measured from the oscillation pole and ranging from 0 to $\pi$, and $\phi$ being the azimuthal angle, ranging from 0 to $2\pi$. Each of the grid points has a local effective temperature $T_{\mathrm{eff}}$ and a local oscillation velocity vector $\vec{v_{\mathrm{osc}}}$ defined by the three components \citep{Kurtz2006, Hatzes1996, Schrijvers1997, Kochukhov2004}

\begin{align}
    v_r &= \frac{{\partial{\xi_r}}}{\partial{t}} = i 2 \pi \nu  \frac{a}{C} Y_l^m(\theta, \phi)\exp(i2\pi\nu t), \label{Eq:vr}\\
    v_\theta &= \frac{\partial{\xi_\theta}}{\partial{t}} = i 2 \pi \nu  \frac{Ka}{C} \frac{\partial Y_l^m(\theta, \phi)}{\partial\theta} \exp(i2\pi\nu t),\label{Eq:vtheta}\\
    v_\phi &= \frac{\partial{\xi_\phi}}{\partial{t}} = i 2 \pi \nu \frac{Ka}{C\sin\theta} \frac{\partial Y_l^m(\theta, \phi)}{\partial\phi})\exp(i2\pi\nu t)\label{Eq:vphi}, 
\end{align}
which are oriented in the directions of the local unit vectors of r, $\theta$, and $\phi$, respectively. We use the real part of the velocity components to calculate the final local velocities. Here, $\Vec{\xi}$ describes the local displacement vector, $\nu$ is the oscillation frequency, $a$ is the displacement amplitude in the radial direction, $K$ is the ratio between the radial and the horizontal displacement, and $Y_l^m$ are the spherical harmonics defined as 
\begin{equation}
    Y_l^m(\theta, \phi) = \sqrt{\frac{2l+1}{4\pi}\frac{(l-m)!}{(l+m)!}}P_l^m(\cos\theta)\exp(im\phi).
\end{equation}
Here we introduce the quantum numbers $l$ and $m$. $l$ is the number of line of nodes on the stellar surface, while $m$ is the azimuthal order quantifying how many of these line of nodes run through the oscillation pole. It ranges from $-l \leq m \leq l$.
$P_l^m(\cos\theta)$ are the Legendre Polynomials defined by
\begin{equation}
    P_l^m(\cos\theta) = \frac{(-1)^m}{2^ll!} (1-\cos^2\theta)^{\frac{m}{2}}\frac{\mathrm{d}^{l+m}}{\mathrm{d}\cos^{l+m}\theta} (\cos^2\theta-1)^l.
\end{equation}
Thus, the velocity amplitude in the radial direction is given by $v_\mathrm{osc} = 2\pi \nu \frac{a}{C}$, and consequently $K \cdot v_\mathrm{osc}$ is the velocity amplitude in the $\theta$ and $\phi$ directions. We introduce the normalization factor $C=\max(\operatorname{Re}(i Y_l^m(\theta, \phi)))$ for $m \neq 0$, and $C=\max(\operatorname{Re}(Y_l^m(\theta, \phi)))$ for $m=0$, such that $v_\mathrm{osc}$ can be regarded as the amplitude of the radial component at the position of maximum variation on the stellar surface, giving it a straight-forward interpretation. Both $v_\mathrm{osc}$ and $K$ are input parameters to be defined by the user. In the Cowling approximation, neglecting perturbations of the gravitational potential, $K$ can be expressed as \citep{Aerts2021, Cowling1941}
\begin{equation}
K = \frac{GM_\star}{\left(2\pi\nu\right)^2R_\star^3}.  
\end{equation}

Associated with the (physical) radial displacement are variations of the local effective temperature of a scale $\delta T_{\mathrm{eff}}$ and at a phase $\psi_T$. Assuming non-adiabatic conditions, we can express the local effective temperature as \citep{Dupret2002, DeRidder2002} 
\begin{equation}
    T_{\mathrm{eff, local}} = T_{\mathrm{eff}} + \delta T_{\mathrm{eff}} \operatorname{Re}\left(\frac{Y_l^m(\theta, \phi)}{\max(\operatorname{Re}(Y_l^m(\theta, \phi)))}\exp(i(2\pi\nu t+\psi_T))\right),
\end{equation}
defining $\delta T_{\mathrm{eff}}$ as the amplitude of the temperature variation at the position of maximum variation on the stellar surface. The scale of the temperature variation is not influenced by the choice of the oscillation velocity $v_\mathrm{osc}$, such that a reasonable value for $\delta T_{\mathrm{eff}}$ has to be chosen by the user.

As the velocity components of interest are generally low, we neglect geometrical distortions of the stellar sphere and keep the stellar radius at unity at all times. 
Thus, the simulations can be visualized as mapping the oscillation's velocity components onto the unit sphere, neglecting the displacements of the individual positions. We further do not include surface-normal variations which were found to play a minor role on the variation of line widths by \citet{DeRidder2002}. However, we note that \citet{Townsend1997} advocates to include both surface-normal and surface-area variations to accurately reproduce photometric continuum variations. 

\subsection{Calculation of the synthetic spectrum}
After choosing an inclination angle, defined as the angle between the polar axis of the oscillation and the line of sight, the velocity and temperature fields are projected onto a flat, uniformly spaced 2D grid of size $N_\mathrm{grid} \times N_\mathrm{grid}$, using triangulation and linear interpolation. The directional velocity components are further projected onto the line-of-sight vector, and can thereafter be summed up to yield the local combined oscillation velocity for each element on the grid. 

We further add a projected rotational velocity for each position characterized by the (unprojected) rotational 
velocity $v_{\mathrm{rot}}$. We fix the axis of rotation to coincide with the polar axis of the simulated star and thus with the symmetry axis of the oscillation. The summation of the projected, combined oscillation velocity and the projected rotational velocity yield the final, local velocities along the line of sight $v_i$ for each position $i$. This local velocity will subsequently be used to Doppler shift the local spectra as detailed below. 

The projected positions on the grid can be regarded as the center points of small surface areas with the same projected area. For cells at the limb of the star, a geometric weight $w_i$ is calculated as these are only partially covering the stellar disk. 

For each grid cell, a synthetic stellar spectrum is calculated by performing a cubic spline interpolation with respect to the local effective temperature $T_{\mathrm{eff},i}$ within a grid of synthetic spectra taken from the PHOENIX spectral library\footnote{\url{https://phoenix.astro.physik.uni-goettingen.de/}} \citep{Husser2013}. For computational speed, the cubic spline interpolation is performed in $0.1\,\mathrm{K}$ steps.  For simplicity, we do not interpolate the spectra with respect to the surface gravity $\log g$ and the metallicity $[\mathrm{Fe/H}]$. Both are fixed to the closest value for NGC 4349 No. 127, that is $\log g=2.0$ and $[\mathrm{Fe/H}]=0.0$. 

Ideally, one would like to use the specific intensity spectra available from the PHOENIX library, which were calculated for a model atmosphere under different observation angles $\mu = \cos{\gamma}$, with $\gamma$ being the angle between the line of sight and the surface normal. However, these are only available with a sampling rate of $1.0\,\mathrm{\AA}$, too low to simulate high-resolution Echelle spectra. We are therefore forced to use the high-resolution spectra of the PHOENIX spectral library. As these are already disk-integrated spectra, the final simulated spectrum can only be regarded as a relatively crude approximation. 

\subsubsection{Implementation of the limb-darkening correction}
Since we are interested in the observational effects of non-radial oscillations, which can be dominated by their horizontal components and thus strongest toward the limb of the stellar disk, we implement a wavelength-dependent limb-darkening correction. \citet{Hestroffer1998} fit the intensity profile of the solar disk as a function of $\mu$ and wavelength $\lambda$ using data provided by \citet{Pierce1977} and \citet{Neckel1994} and provide the relation 
\begin{equation}
    I(\mu, \lambda) = 1 - u(1-\mu^{\alpha(\lambda)}).
\end{equation}
\citet{Hestroffer1998} use $u=1$ and determine $\alpha \sim -0.023 + 0.292 \lambda^{-1}$ for $\lambda$ in units of $\mathrm{\mu m}$ and $416\,\mathrm{nm} \lesssim \lambda \lesssim 1099\,\mathrm{nm}$. Since the wavelength range of HARPS extends down to $378\,\mathrm{nm}$ at the blue end, we extrapolate the model, despite the discontinuity at $\lambda \sim 390\,\mathrm{nm}$ discussed by \citet{Hestroffer1998}.

As the underlying PHOENIX spectra are already disk-integrated and therefore include the effect of limb darkening, we first divide each local spectrum by the wavelength-dependent intensity profile for the mean angle $\Bar{\mu}$. Next, the limb-darkening effect is added back into the models by multiplying with the respective intensity profile for each cell according to its local angle $\mu_i$. By this approach, we add an implicit weight based on the intensity according to the angle $\mu_i$ during the spectrum combination process detailed below. Ideally, one would like to use $\mu$ and wavelength-dependent models for a star more similar to NGC 4349 No. 127, but such models are not available in the literature.

We further include the effect of macroturbulent broadening by convolving the local spectra with a wavelength-dependent Gaussian kernel, specified by a user-defined macroturbulent velocity $\zeta$. Microturbulent broadening has already been included during the synthesis of the base PHOENIX spectrum, albeit at a fixed value for each grid point within the PHOENIX library \citep{Husser2013}.

\subsubsection{Implementation of the convective blueshift correction}
We also include an estimation for the effect of convective blueshift on the spectral line bisectors, following the approach by \citet{Zhao2023}. As the PHOENIX spectral library is derived from 1D atmospheric models, they do not properly reproduce line bisector shapes \citep{Zhao2023}. The intrinsic bisectors of the PHOENIX spectra therefore have to be removed prior to adding in a more plausible bisector shape stemming from convective processes.

First, the PHOENIX spectra are normalized using RASSINE \citep{Cretignier2020}. Using the normalized spectra we first calculate the small wavelength shifts necessary to remove the intrinsic PHOENIX bisector by measuring the bisectors of five FeI lines (FeI $5250.2084\,\textrm{\AA}$, FeI $5250.6453\,\textrm{\AA}$, FeI $5434.5232\,\textrm{\AA}$, FeI $6173.3344\,\textrm{\AA}$, FeI $6301.5008\,\textrm{\AA}$) and calculating the average. Next, the wavelength shifts from one of the giant stars presented by \citet{Gray2005} are calculated using the bisector-removed, normalized spectra. Both the wavelength shifts are applied to each grid point in the unnormalized PHOENIX spectra. These are then used during the spectrum combination process. By testing all available bisectors, we found that the observed CCF bisectors are well reproduced using the bisector model of the giant star $\beta$ Boo. While this procedure creates CCF bisectors quite similar to the observed ones, it has little effect on the observables presented in Sect.~\ref{sec:results}.

\subsubsection{Combination of the local spectra}
Finally, after including each of the effects on the local spectra (local temperature variation, limb darkening, convective blueshift), the individual spectra have to be Doppler-shifted before being summed to yield the disk-integrated spectrum.

To yield accurate Doppler shifts, we first oversample the wavelength grid at equidistant $0.001\,\textrm{\AA}$ steps and perform a cubic spline interpolation for the flux values at the oversampled grid points. After the spectra are Doppler-shifted according to the summed local velocities of oscillation and rotation $v_i$, the shifted flux values are then linearly interpolated back onto the original wavelength grid, which is common for all surface grid points. Finally, the spectra are summed, accounting for the reduced geometrical weights $w_i$ for cells at the edge of the star, and divided by the sum of all geometrical weights to maintain a reasonable normalization of the spectra. The final, disk-integrated spectrum can be expressed as 
\begin{equation}
    I_\mathrm{integrated}(\lambda) = \frac{1}{\sum_i w_i} \sum_i I_i(T_{\mathrm{eff},i}, v_i, \mu_i, \lambda) \cdot w_i .
\end{equation}

Applying the above steps, we calculate synthetic stellar spectra including the effect of non-radial oscillations. The code is also able to study other effects such as star spots, which is however beyond the scope of this publication.

\subsubsection{Conversion to simulated HARPS spectra}

The combined spectrum is next smoothed with a Gaussian kernel to bring the spectrum to the required resolution for each spectrograph. For HARPS, $R = 115\,000$ \citep{Mayor2003} is used (CARMENES VIS and NIR channels are also available). The smoothed spectra are then rebinned onto an extracted wavelength grid of a real HARPS observation of NGC 4349 No. 127. Conceptually, converting the combined spectrum from an energy flux to a photon flux would more closely mimic the acquisition process of a real CCD. However, tests revealed that the differences are negligible and we thus decided to omit the conversion for simplicity. Next, a measured blaze function is applied to each spectral order and the necessary FITS header keywords are altered to mimic a real observation. Finally, the spectrum is saved in the FITS format and the RVs can be reduced from the simulated, extracted spectra. The simulation is performed at user-defined epochs. 

The final spectra are then reduced using the RACCOON and SERVAL pipelines -- identical to the processing applied to real observations -- yielding the simulated RVs and activity indicators. For the RACCOON pipeline, we use the weighted mask created from the real observations. For the SERVAL reduction, a new template is created for the simulation. 

For the simulation results presented below, we performed the simulation with a grid of size $150 \times 150$. This number was chosen for computational speed, but tests at higher resolutions revealed negligible differences.

\section{Results}
\label{sec:results}
\subsection{Confirmation of the intrinsic nature of the RV variations}
From the 46 spectra acquired prior to the HARPS fiber change in 2015, two (BJD = 2454323.471811, BJD = 2454349.472032) were found to be strong outliers in the dLW time series. These have $\mathrm{dLW} = 96.6$ and $\mathrm{dLW} = 96.4$, respectively, while the dLW time series for the other spectra varies between $-21$ and $24$. These two spectra were discarded from all time series. We note that the two data points do not have particularly low S/N and are not conspicuous in the RVs or any of the other activity indicators. It remains unclear, why the two spectra are strong outliers in only one indicator.

Another strong outlier (BJD = 2458849.842858) in the RV time series was removed from the 12 spectra acquired after 2015. Again, it remains uncertain why the spectrum deviates strongly in the RVs. It has sufficient S/N, albeit slightly below average.

\begin{figure}
    \centering
    \includegraphics{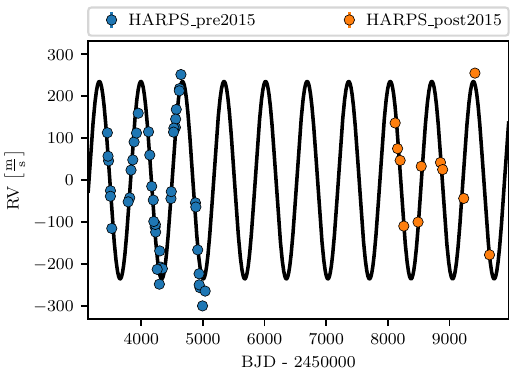}
    \caption{RVs reduced by SERVAL prior to (blue) and after (orange) the HARPS fiber change in 2015 plotted against time. The error bars are smaller than the size of the markers. A sinusoidal fit is plotted in black and reveals the RVs to be consistent with a long-lived, coherent signal that could (when examined in isolation to other diagnostics) be attributed to a brown dwarf orbiting the primary.}
    \label{fig:rv}
\end{figure}

The remaining 44 spectra taken before the fiber change and 11 spectra taken after the fiber change are plotted in Fig.~\ref{fig:rv} as blue and orange data points, respectively. We treated the two data sets independently and fitted for an offset.
We plot a fitted sinusoid with period $P=674.0\pm0.1\,\mathrm{d}$ in black to illustrate that the RV signal is consistent with a periodic variation stable over at least fifteen years. This RV variation by itself could easily be attributed to an orbiting companion, which was however shown not to exist by \citet{DelgadoMena2018, DelgadoMena2023}, based on significant periodicity of the FWHM of the CCF close to the proposed orbital period. \citet{DelgadoMena2018} further report periodicity of the $\mathrm{H\alpha}$ index close to the orbital period but with a false alarm probability (FAP) less significant than $1.0\%$, which is further reduced in significance when including the spectra acquired after 2015 \citep{DelgadoMena2023}. Moreover, the authors find a weak but significant correlation between $\mathrm{H\alpha}$ and RV.

\begin{figure}
    \centering
    \includegraphics{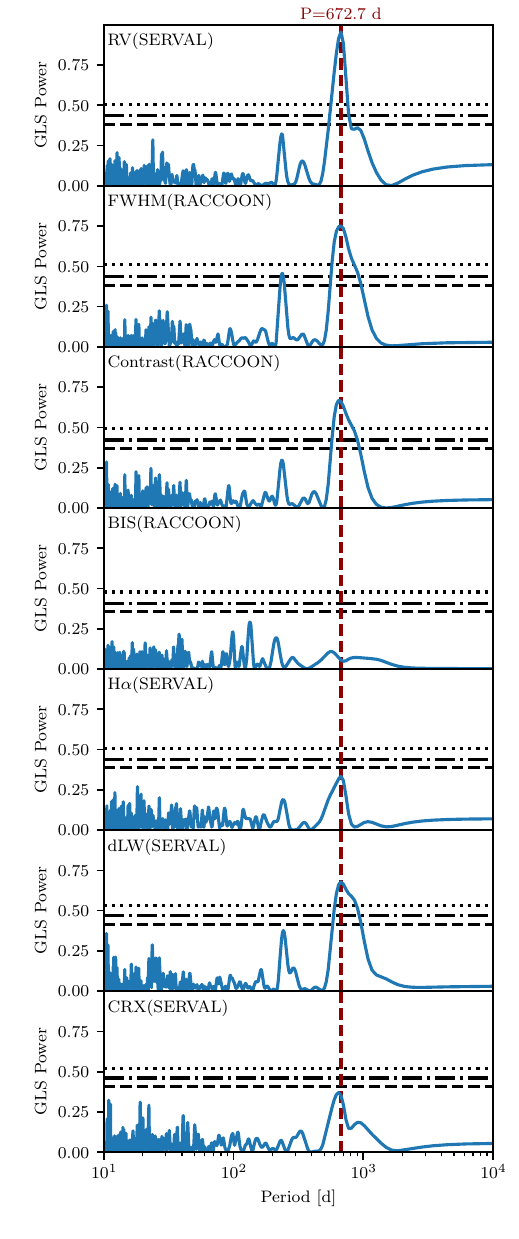}
    \caption{GLS periodograms of the RVs and activity indicators calculated for the 44 HARPS spectra acquired prior to the fiber change. The FAP of 5\% (dashed line), 1\% (dash-dotted line), and 0.1\% (dotted line) were determined using a bootstrap with 10\,000 reshuffles and are plotted for each panel. The strong RV periodicity at $P=672.7\,\mathrm{d}$ is accompanied by significant periodicity of the FWHM and contrast of the CCF and the dLW. The $\mathrm{H\alpha}$ indicator of the SERVAL reduction and the CRX show peaks at or very close to the orbital period that, however, have FAP > 5\%. The BIS is inconspicuous.}
    \label{fig:periodogram}
\end{figure}

With the additional activity indicators available through the SERVAL and RACCOON reductions, we can strengthen these findings. Figure~\ref{fig:periodogram} shows a generalized Lomb-Scargle (GLS) periodogram \citep{Zechmeister2009} of the 44 spectra acquired prior to the HARPS fiber change and reveals that the RV periodicity in the pre 2015 data set (red dashed line) is accompanied by further significant peaks of the activity indicators close to the orbital period. Besides the FWHM, also the contrast of the CCF and the dLW of the SERVAL reduction show strong peaks more significant than the $\mathrm{FAP}=0.1\%$ level, which was determined via bootstrapping with 10\,000 reshuffles. The peak in the contrast GLS is slightly offset at $P=645.4\,\mathrm{d}$. 

Both the $\mathrm{H\alpha}$- and the CRX-periodograms furthermore have peaks at the orbital period, or slightly offset in case of the CRX ($P = 645.4\,\mathrm{d}$). While both peaks are formally less significant than the $\mathrm{FAP}=5\%$ level, the fact that they appear very close to the RV period and are the largest peak in their respective periodograms is certainly not a coincidence and indicates a signal at the RV period in H$\alpha$ and CRX.
Only the BIS periodogram is inconspicuous. Taken together, the additional spectral diagnostics strengthen the findings by \citet{DelgadoMena2018, DelgadoMena2023}, rule out a physical companion, and thus confirm the intrinsic origin of the RV variations.

\subsection{Correlations between the activity indicators and the RVs}

\begin{figure*}[ht]
    \centering
    \includegraphics{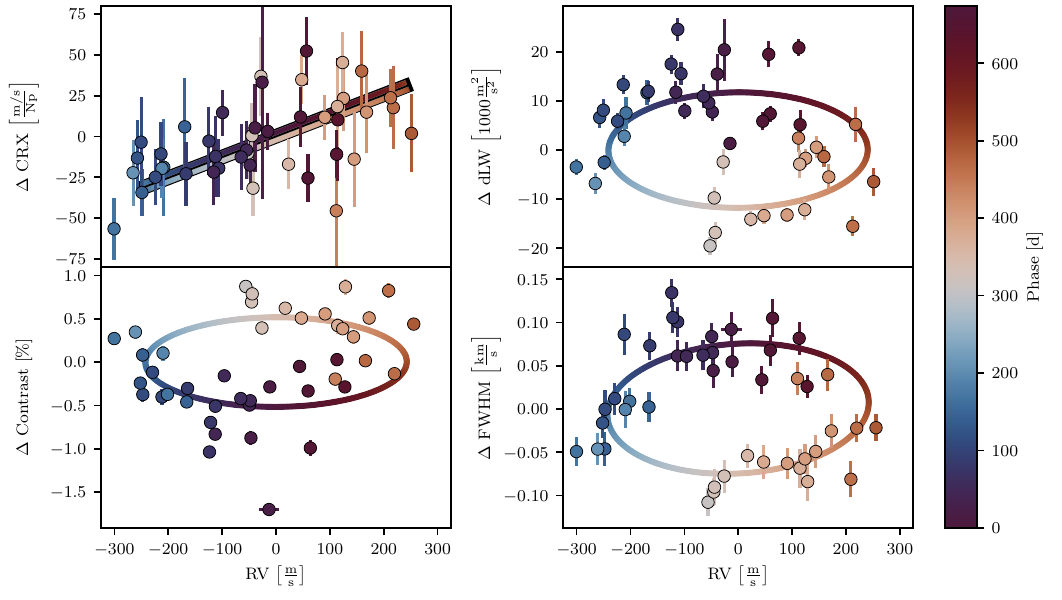}
    \caption{Observed (data points) and simulated (lines) correlations between the activity indicators and the RVs for 44 spectra taken prior to the HARPS fiber change. For each panel, the mean (RV and any of the indicators, respectively) was subtracted.
    Each data point is color-coded with the phase according to the best RV period $P=674.0\,\mathrm{d}$. 
    While the CRX (top left) shows a significant positive correlation with the RVs (r=0.54, $p(\textit{F}\textrm{-test})=0.002\%$), dLW, FWHM, and contrast of the CCF are correlated with the RVs in a closed-loop behavior. 
    We plot the linear (CRX) and elliptical fits (dLW, FWHM, contrast) to the simulated data points for the best model of a $l=1, m=1$ oscillation mode as solid lines applying the same color-coding. As a linear relationship is predicted between CRX and RV, we plot the ascending and descending phase relations on top of the black fit to the simulations. The simulated ellipses can closely reproduce the observed behavior including the amplitudes, phases, and directions of correlation.}
    \label{fig:corr}
\end{figure*}

\begin{figure*}[h!]
    \centering
    \includegraphics{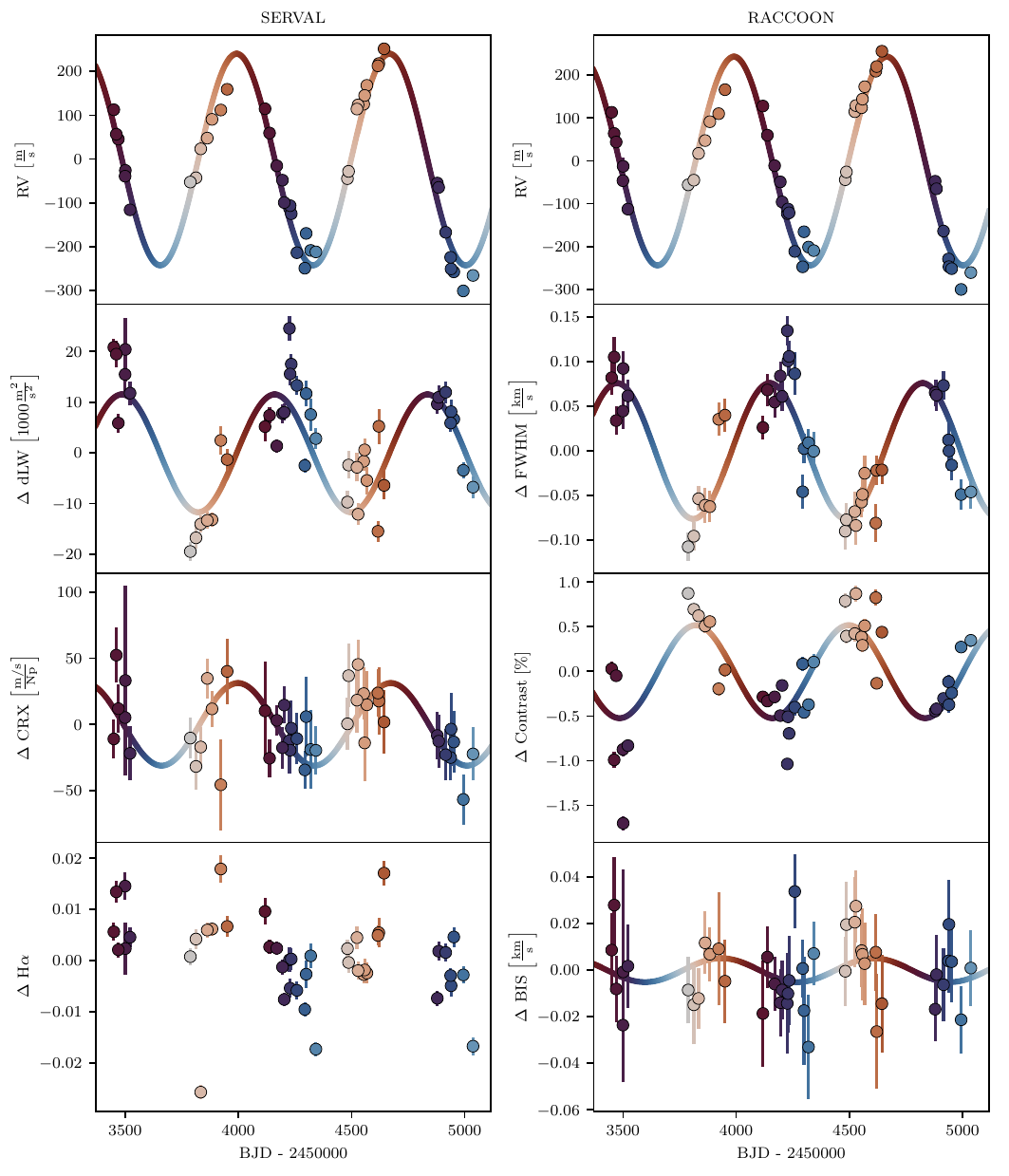}
    \caption{Observed (data points) and simulated (lines) RVs and activity indicators of the 44 HARPS spectra acquired prior to 2015 plotted against time. The same color-coding as in Fig.~\ref{fig:corr} was applied. The mean RV was subtracted from both the SERVAL and RACCOON RV time series, as well as the mean of each indicator time series The color-coded solid lines are sinusoidal fits to the simulated spectra computed for the best $l=1, m=1$ mode. The simulation is able to reproduce the amplitudes and phases of all indicators, although some (real) indicators suffer from large scatter. As the simulations focus solely on the stellar photosphere, no meaningful variation of the $\mathrm{H\alpha}$ indicator can be simulated. }
    \label{fig:timeseries}
\end{figure*}

To find an alternative explanation for the radial velocity variations, we searched for correlations between the activity indicators and the RVs. 
Figure~\ref{fig:corr} shows the CRX, dLW, contrast, and FWHM of the CCF plotted against the RV for the 44 HARPS spectra taken prior to the 2015 fiber exchange. For all indicators, we plot the absolute deviation from the mean value, which was determined by fitting sinusoids and determining the offset, to allow an easier comparison with the simulated variations.  

All data points in Fig.~\ref{fig:corr} are color-coded with the RV period $P=674.0\,\mathrm{d}$ present in the full data set (including the spectra taken after 2015). It is evident from the top left panel that the CRX is positively correlated with the RV with a Pearson's r value $r=0.54$ with p-value $p(r)=0.015\%$. We fitted a linear relation using orthogonal distance regression and perform an \textit{F}-test \citep{Fisher1925} to validate the relation's significance against a constant model, resulting in a p-value $p(\textit{F}\textrm{-test})=0.002\%$. Following \citet{Tal-Or2018} and \citet{Benjamin2018}, we regard the correlation as significant. As \citet{DelgadoMena2023}, we further note a significant positive correlation of the $\mathrm{H\alpha}$ index with the RV with $r=0.48$ and $p(\textit{F}\textrm{-test})=0.2\%$ (shown in Fig.~\ref{fig:halpha}).

Furthermore, we observe closed-loop correlations with the RV for the dLW, contrast, and FWHM of the CCF. That is, the activity indicators correlate elliptically with the RVs with each data point's position on the ellipse given by the phase of the variation, indicated by the color-coding. Such ellipses are effectively Lissajous curves resulting from sinusoidal variations at the same period but with different amplitudes and phase shifts close to $\frac{\pi}{2}$.

The color-coding further reveals the direction of correlation, which is anti-clockwise for FWHM and dLW but clockwise for the contrast, which is to be expected as FWHM and contrast should be inversely dependent on each other. We note that the dLW measures a similar line variation as the FWHM and the contrast and should therefore not be regarded as an entirely independent indicator \citep{Zechmeister2018, Jeffers2022}. However, as the results stem from different reduction pipelines (RACCOON vs SERVAL) with different approaches to derive the RVs, it is reassuring that we find a similar behavior for both reductions.  

\subsection{Simulation of a dipole, retrograde (\texorpdfstring{$l=1, m=1$}{l=1, m=1}) oscillation mode}

In order to explain these peculiar correlations, we simulated a set of models using the \texttt{pyoscillot} simulation suite detailed in Sect.~\ref{sec:simulations}. 
Motivated by oscillatory convective modes as published by \citet{Saio2015}, we tested different configurations for dipole ($l=1$) modes using the stellar parameters ($T_{\mathrm{eff}}=4500\,\mathrm{K}, \log g = 2.0, [\mathrm{Fe/H}]=0.0,\zeta=4822\,\mathrm{m\,s^{-1}}$) for the base spectrum as discussed in Sect.~\ref{sec:observations}. 

We include the chromatic effect of limb darkening and the effect of convective blueshift. For the latter, we calculated the CCF bisectors for all measured bisectors of giant stars from \citet{Gray2005} and compared them to the observed CCF bisectors for NGC 4349 No. 127. The best match was found for the bright giant star $\beta$ Boo. We note, however, that the inclusion of the convective blueshift plays a minor role in modeling the relative variations of the activity indicators and mostly provides an offset for the absolute value of the BIS. 
\begin{figure}
    \centering
    \includegraphics{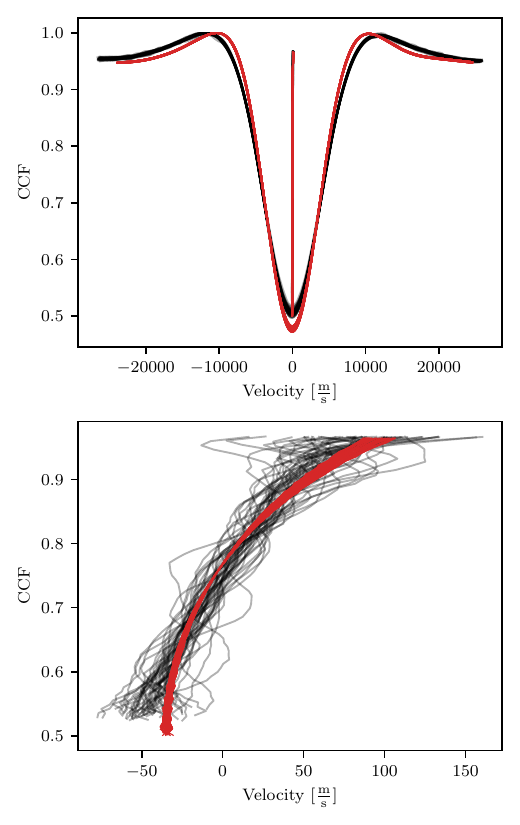}
    \caption{CCFs and bisector profiles for the observed spectra (black) and the simulated spectra (red). Top: CCFs plotted against the velocity shift for the observed and simulated spectra. The CCFs were shifted with the RV to lie on top of each other. The measured bisector for each line is overplotted.
    Bottom: Zoomed-in image of the measured bisectors. The simulated bisectors (red) show much less scatter than the observed ones, but generally follow the same rightward trend. }
    \label{fig:bis}
\end{figure}

Using the RV period and the stellar properties, the ratio between the horizontal and radial oscillation components can be estimated as $K=1856\pm404$. Therefore, the oscillation is dominated by the horizontal components, such that the product $K \cdot v_\mathrm{osc}$ is the decisive input parameter and should be interpreted in conjunction (since changes in $v_\mathrm{osc}$ can be balanced out by changing $K$). This behavior was confirmed through test simulations. $K$ was fixed to $K=1856$ when searching for the best fitting parameters. 

Initial tests at different inclination angles revealed that retrograde ($m=1$) modes are the only dipole modes capable of reproducing the observed correlations for NGC 4349 No. 127, as the other allowed modes would either lead to much smaller amplitudes in the indicators ($m=0$) or to inverse phase relations ($m=-1$). We examine this behavior more closely in Sect.~\ref{sec:results:l=1modes} and focus on $l=1, m=1$ modes for the current discussion. We examine the amplitudes for inclination $i=90\degr$ (equator-on), which maximizes the amplitude for the $|m| = 1$ modes. We also restrict the models to only include one isolated oscillation mode. 

We further find that the scale of the temperature variations $\delta T_\mathrm{eff}$ directly influences the amplitudes of all activity indicators and the RV. Nevertheless, as the CRX captures the wavelength dependence of the RV, it is the most sensitive indicator to the $\delta T_\mathrm{eff}$ parameter. At the same time, the variations of the CRX are strongly influenced by the chromatic limb-darkening law. At maximum RV, the visible and projected $\theta$ and $\phi$ components of the $l=1, m=1$ oscillation are entirely positive. As the RVs are dominated by the horizontal components, the strongest contribution stems from the cells at the limb of the star due to the projection onto the line of sight. The limb darkening suppresses their contribution, such that the overall RV is reduced compared to a model without limb-darkening correction. However, as the limb-darkening law is chromatic, this reduction is stronger at shorter wavelengths, leading to a larger RV at longer wavelength and hence a positive CRX at positive RV. This accounts for the majority of the positive CRX-RV correlation.

Only small temperature variations are therefore consistent with the data, as an increase in $\delta T_\mathrm{eff}$ at phase shift $\psi_T=0$ increases the CRX-RV slope. However, at such small scales the scatter within the data complicates an exact determination. We therefore estimated a value $\delta T_{\mathrm{eff}}=2.5\,\mathrm{K}$ predicted for an oscillatory convective mode as presented in Fig.~6 of \citet{Saio2015}. We determined this value based on the luminosity of the star and the radial amplitude that minimized the residuals between the simulations and the real data for the other observables. We note that the models presented by \citet{Saio2015} at the luminosity of NGC 4349 No. 127 are only available for lower stellar masses. A higher temperature variation is therefore possible. 

For small temperature variations, the simulations become rather insensitive to small deviations of the phase shift between the temperature variation and the radial displacement, given the scatter in the real data. This phase shift is predicted to be small, $\psi_T \lesssim 0.1 \cdot \pi$ \citep{Saio2015}. We therefore fixed this value to $\psi_T=0$, such that the temperature is largest when the radial displacement is at its maximum.

With these initial considerations, we ran a large number of models varying $v_{\mathrm{rot}}$ and $v_\mathrm{osc}$, finally selecting the model (by eye) that minimizes the residuals between the simulations and the real data for all observables. 
We simulated 40 synthetic spectra spread over the time span of the real observations (prior to the fiber change) and reduced these using SERVAL and RACCOON. We find that the amplitudes of the activity indicators and the RVs and their phase relations are reproduced using $v_{\mathrm{rot}}=1700\,\mathrm{m\,s^{-1}}$ and $v_\mathrm{osc}=0.30\,\mathrm{m\,s^{-1}}$. We summarize the best values for all simulation parameters in Table~\ref{tab:oscillationparams}.

\begin{table}
    \centering
    \caption{Parameters used for the retrograde dipole oscillation model.}
    \label{tab:oscillationparams}
    \begin{tabular}{ll}
    \hline
         Parameter & Value\\
         \hline
         
         $T_{\mathrm{eff}}$ & $4500\,\mathrm{K}$  \\
         $\log g[\mathrm{cm\,s^{-2}}]$ & $2.0$ \\
         $[\mathrm{Fe/H}]$ & $0.0$ \\
         $l$ & 1 \\
         $m$ & 1 \\
         $P$ & $674.0\,\mathrm{d}$ \\
         $v_\mathrm{osc}$ & $0.30\,\mathrm{m\,s^{-1}}$ \\
         $K$ & $1856$ \\
         $v_{\mathrm{rot}}$ & $1700\,\mathrm{m\,s^{-1}}$ \\
         $\zeta$ & $4822\,\mathrm{m\,s^{-1}}$ \\
         $\delta T_{\mathrm{eff}}$ & $2.5\,\mathrm{K}$ \\
         $\psi_T$ & $0\degr$ \\
         $i$ & $90\degr$ \\
         $N_{\mathrm{grid}}$ & $150\times150$ \\
         \hline
    \end{tabular}
\end{table}

\subsection{Comparison between the data and the oscillation model}
Equipped with the non-radial oscillation model, we compare the observed correlations with the simulated ones in Fig.~\ref{fig:corr}. The aim is to test whether such an oscillation is able to explain the observed behavior.

 The oscillation model predicts a positive correlation between CRX and RV (top left panel). We fitted the correlation with a linear relation and determined the slope $m_\mathrm{sim}=0.129\pm0.001\,\mathrm{Np^{-1}}$ (black line). The overplotted, colored lines encode the phase information on the linear relation. The simulated data points deviate only marginally from the linear relation and were left out for visual clarity. The slope is consistent with the actual slope measured for the real data $m_\mathrm{real}=0.101\pm0.021\,\mathrm{Np^{-1}}$ with a deviation of $1.3\,\sigma$.

For the dLW, contrast, and FWHM, closed-loop correlations very similar to the observed correlations are predicted. We plot fitted ellipses to the simulated indicator-RV correlations that closely follow the synthetic data points (which were therefore left out for visual clarity) in Fig.~\ref{fig:corr}, applying the same color-coding. An offset for the RV zero points was fitted and removed between the real and synthetic data sets. We stress again that we show the absolute deviation from the respective mean of each time series. The absolute values between the FWHM and contrast of the CCF of the real and synthetic observations are slightly offset as not all influences can be modeled adequately. 

We also had to downscale the simulated dLW variations by a multiplicative factor of 0.098 to match the real dLW variations. As the dLW is a differential quantity sensitive to the second derivative of the SERVAL templates, which are separate for the real and simulated data sets, such a multiplicative factor is expected when comparing separate dLW time series \citep{Zechmeister2018}. We  find that the scaling factor is on the order of unity when using the same template for both reductions, which however reduces the precision of individual RV and indicator determinations. 
The multiplicative factor is presumably caused by the (slightly) different widths of the real and simulated spectral lines (see also Sect.~\ref{sec:cross-correlation profiles}). It is furthermore linked to deviations from the assumption of Gaussian spectral lines due to the oscillation, which are fundamental to the definition of the dLW \citep{Zechmeister2018}. We note that, due to this scaling factor, the amplitude of the dLW holds little quantitative information. The phase relation with the RV, however, holds valuable information to validate the simulations.       

It can be seen from Fig.~\ref{fig:corr} that all variations of the activity indicators, the relative phases, and the directions of correlations are well reproduced. We tested whether radial, p-mode (solar-like) oscillations can explain the residual scatter present in the activity indicators. From the residual scatter in the RVs, we calculated associated temperature variations of $\delta T_{\mathrm{eff}} \sim 0.9\,\mathrm{K}$ using the scaling relations by \citet{Kjeldsen1995}. Fine-tuning $v_\mathrm{osc}$ to reproduce the residual scatter in the RVs, we find that the activity indicators are in principle sensitive to the p-mode oscillations, but the amplitudes are too small to explain the observed residuals. We assume that instrumental effects and other intrinsic, stellar noise sources, such as granulation, dominate the residual scatter. 

The CRX is sensitive to the wavelength dependence of the RVs and thus to temperature variations on the stellar surface, as well as the limb-darkening coefficients. We find the interplay of limb darkening and the oscillations to be the dominating influence. As the limb-darkening contrast decreases toward redder wavelength and the resulting RV is dominated by the horizontal components at the limb of the star, the resulting RV amplitude is larger in the redder part of the spectrum and hence the CRX is positively correlated with the RV. Its slope is commonly referred to as chromaticity \citep{Zechmeister2018}. 
  
In comparison, the temperature variations have only minor influence on the observed CRX-RV slope. If we simulate an (unphysical) oscillation model setting $\delta T_\mathrm{eff}=0\,\mathrm{K}$, the resulting slope of the CRX-RV correlation is reduced slightly to $m_{\delta T=0K}=0.1213\pm0.0003\,\mathrm{Np^{-1}}$, even closer to the real correlation. A limb-darkening model adapted to the stellar parameters of NGC 4349 No. 127 could therefore be beneficial, but is beyond the scope of this work. 
 
The variations of the line shape indicators (dLW, FWHM, contrast) are mainly influenced by the interplay between the oscillation at the edge of the stellar disk (for which the dominating horizontal components are most directly oriented along the line of sight) and the rotational broadening. This can be easily understood as the oscillation velocities at the limb directly affect the flanks of the broadened spectral lines, thus introducing variations of the line width and depth, or introducing asymmetries. 

Figure~\ref{fig:timeseries} shows the RV and indicator time series data acquired pre 2015. We fitted sinusoids to the simulated time series and color-code these in the same way as the real data. The simulations are able to reproduce the amplitudes and phases of the RVs and all activity indicators, although large scatter is present in some of the real indicators. We note that the simulations also predict small variations of the BIS at the same period as the RVs that are consistent with the observed data, thus adding another argument in favor of the oscillation model. Large errors and high scatter in the real data likely obscure these variations, so that they are not detected in the GLS periodogram search in Fig.~\ref{fig:periodogram}.

No meaningful $\mathrm{H\alpha}$ variations can be predicted by the models since the simulations focus solely on the stellar photosphere, while the $\mathrm{H\alpha}$ variations are mostly caused by chromospheric processes \citep{Kurster2003}. Whether such chromospheric (and likely magnetic) processes could be linked to the oscillations is unknown. We note that similar connections between stellar oscillations and magnetic fields have been proposed (see, e.g., \citealt{Lebre2014, Georgiev2023, Konstantinova-Antova2024}).

\subsection{Cross-correlation profiles}
\label{sec:cross-correlation profiles}

In Fig.~\ref{fig:bis} (top), we plot the real, observed CCF profiles from the RACCOON reduction for NGC 4349 No. 127 (black) as well as the modeled profiles (red). While the CCF shapes are reasonably well reproduced, slight differences in the absolute values of the CCF width (FWHM) and its contrast remain. We therefore plot the differences from the respective means in Fig.~\ref{fig:corr} and Fig.~\ref{fig:timeseries}. 

Offsets between the simulated and the real CCF profiles are to be expected as not all influences can be modeled accurately. These include the real instrumental profile of the HARPS spectrograph, which we model as a Gaussian, and the intrinsic line shapes inherent to the PHOENIX models. These are given with fixed microturbulent broadening values for each model. For the base PHOENIX model employed in the simulation, microturbulence $\xi = 1.49\,\mathrm{km}\,\mathrm{s^{-1}}$ was used, somewhat lower than expected for our star (see Table~\ref{tab:params}). Due to these inherent differences, the line shapes cannot be reproduced accurately. 

We also find that the rotational velocity $v_{\mathrm{rot}}$ plays an important role in reproducing the amplitudes of the line shape indicators (dLW, FWHM, contrast), with higher rotation velocities generally producing larger variations. At the same time, $v_{\mathrm{rot}}$ influences the absolute width of the CCF profile. We find a smaller rotation velocity ($v_{\mathrm{rot}} = 1.7~\mathrm{km}\,\mathrm{s^{-1}}$) to be generally in good agreement with the observed variations for NGC 4349 No. 127. We note that more recent determinations of the star's rotation velocity are larger (see Table~\ref{tab:params}), but find good agreement with the rotation velocity determined by \citet{Carlberg2016}. However, due to the inherent differences of the PHOENIX spectra to the real spectrum of the star, we caution against using this value as a determination of the rotation velocity of NGC 4349 No. 127. 

The bottom panel of Fig.~\ref{fig:bis}  portrays a zoomed-in image of the calculated bisector profiles. We observe that the general bisector shape is well reproduced by the convective blueshift model based on $\beta$ Boo \citep{Gray2005}. However, we find the bisector shape to have negligible effect on the observables presented in Fig.~\ref{fig:corr} and Fig.~\ref{fig:timeseries}, apart from providing an offset for the absolute value of the BIS.

\subsection{Photometric variations}
For many types of intrinsic variations manifested in RV data, such as stellar spots or radial oscillations, photometric variability is expected and can provide independent constraints on their properties (see, e.g., \citealt{Hojjatpanah2020}).
One V-band photometric data set, already discussed by \citet{DelgadoMena2018}, is available for the star from the All Sky Automated Survey (ASAS) at Las Campanas Observatory (Chile) \citep{Pojmanski2004}. We use only grade A and B results as suggested by \citet{DelgadoMena2023}. 

In Fig.~\ref{fig:phot} (top), we plot the ASAS-3 data (dots) against time. We overplot binned data points (black rectangles) with bin size $67.4\,\mathrm{d}$ ($10\%$ of the RV period) to add some visual clarity. There is significant scatter at the $\sim$ 0.1 mag level, albeit without obvious periodicity. We overplot the predicted V-band variations from the simulated dipole oscillation caused by the temperature fluctuations. These were derived by integrating the product of the simulated spectra and the Bessel V-band filter curve\footnote{\url{http://spiff.rit.edu/classes/phys440/lectures/filters/bess-v.pass}} and rescaling to the median magnitude present in the ASAS-3 data set. The simulations predict a sinusoidal variation at the RV period with an amplitude of 0.003 mag. This lies well below the average ASAS-3 error for NGC 4349 No. 127 of 0.051 mag and below the scatter present in the ASAS-3 data set. It would therefore be plausible that the photometric variation stemming from the oscillation would not be detected by the ASAS-3 photometry. We thus conclude that the available photometry does not argue against the oscillation hypothesis. We discuss the available photometry in more detail and comment on the findings by \citet{DelgadoMena2018} in Appendix \ref{sec:app:photometry}.

ASAS-3 is reported to achieve a differential accuracy of 0.01 mag for ideal, bright targets \citep{Pojmanski2004}, which is still larger by a factor of 3 than the photometric amplitude predicted by the simulation. The star was further observed in the TESS full-frame images in sectors 11, 37, 38, 64, and 65. However, the short duration of each individual sector, the long gaps between the available sectors, and instrumental offsets between them prevent us from analyzing the TESS photometry on timescales similar to the RV period. Future {\it Gaia} data releases might provide the necessary photometric precision and timescales to detect such variability.\footnote{\url{https://www.cosmos.esa.int/web/gaia/science-performance}} However, as the star is relatively bright, systematic effects might hinder this analysis.

The calculation of the photometric variability is based solely on the temperature variations on the stellar surface, neglecting surface-area and surface-normal variations. The latter are small at the velocity amplitudes considered but would violate the assumption of spherical symmetry of the star, complicating the calculation of the velocity fields. Given the small amplitudes, we find this simplification justified. 

We note, however, that \citet{Townsend1997} argues to include surface-area and surface-normal variations when predicting photometric variations of non-radial oscillations. However, the oscillation amplitudes considered are much larger than in the case of NGC 4349 No. 127 and their $K$ values range between 0 and 1. The differences between simulations considering these geometrical surface variations and those that do not (see their Fig.~6) seem to become less pronounced toward larger $K$ (as in our case). The surface-normal variations mainly act to avoid an unphysical photometric minimum at $K \sim 0.85$ with the differences being on the order of a factor 2 to 3 otherwise. It can therefore be assumed that the overall photometric variability would not be affected enough to be easily detectable by the ASAS-3 photometry. 

With the presented results, we show that a retrograde, dipole ($l=1, m=1$) oscillation mode is fully consistent with the radial velocity variations and the variations of all studied activity indicators, including the amplitudes and phase relations, as well as the available photometry. We therefore conclude that non-radial oscillations are indeed present in the star and cause the observed periodic patterns.

\section{Other oscillation modes}
\label{sec:othermodes}
Having established that a retrograde $l=1$ mode is able to reproduce the observables, we further aim to qualitatively address the question whether, alternatively, other oscillation modes are also capable to reproduce the same behavior.
\subsection{l=1 modes}
\label{sec:results:l=1modes}
\begin{figure*}
    \sidecaption
    \includegraphics{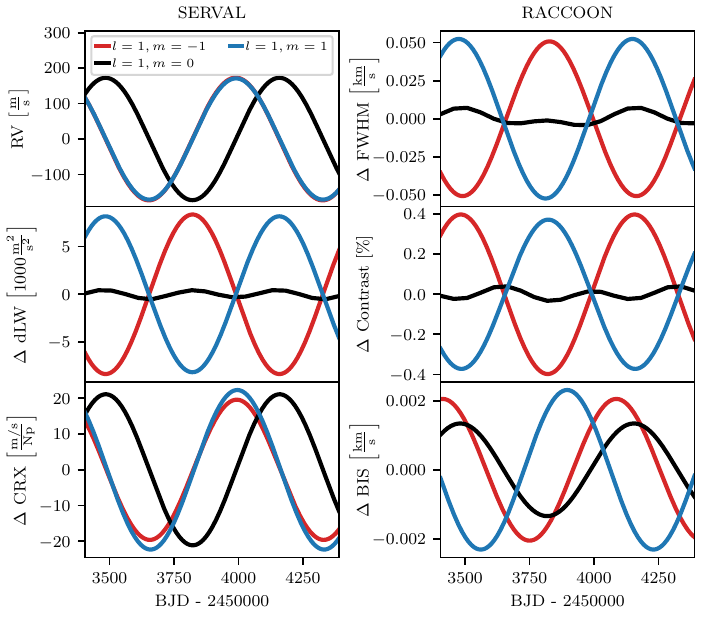}
    \caption{RVs and activity indicators of dipole oscillation modes plotted against time at inclination angle $i=45\degr$. Sinusoids were fitted to all time series except for the FWHM, dLW, and contrast in case of $m=0$, for which the simulated points were interconnected. While all three modes cause similar RV variations (top left panel), the phase relations between the RVs and the line shape indicators, as well as their amplitudes, are notably different. The CRX (bottom left) shows a similar behavior for all three modes. The dLW variations were rescaled with a common factor of $0.1$.}
    \label{fig:l1_time}
\end{figure*}

\begin{figure*}
\sidecaption
    \includegraphics{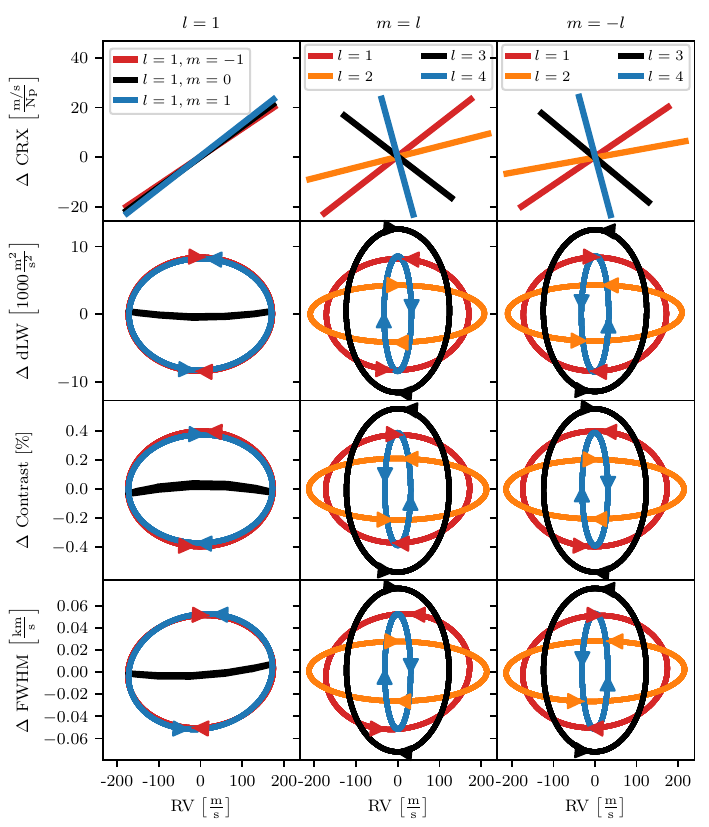}
    \caption{Correlation plots between the activity indicators and the RVs for modes with $l=1$ (left column), $m=l$ (center column), and $m=-l$ (right column). Linear correlations were fitted for the correlation between CRX and RV. For the line shape diagnostics, ellipses were fitted when appropriate. Arrows indicate the temporal dependence of the ellipses. The individual modes are specified in the legend at the top of each column. The simulated dLW variations were rescaled with a common factor of $0.1$.}
    \label{fig:supercorrelation}
\end{figure*}
Motivated by the models for dipole oscillatory convective modes presented by \citet{Saio2015}, we first study the characteristics of $l=1$ modes, for which the azimuthal orders $m=(-1, 0, 1)$ are possible. We chose the same parameters as presented above, aiming to understand the qualitative differences for the different modes. We simulated the different oscillation modes at 20 epochs each. 

We first note that different modes have different inclination angles at which the RV amplitudes are maximized or minimized. In case of the $l=1$ modes, the RV amplitudes are maximized when viewed equator-on ($i=90\degr$) for $|m| = 1$, while the $m=0$ mode has maximum RV amplitude when viewed pole-on ($i=0\degr$) \citep{Chadid2001, DeRidder2002}. We therefore ran the simulations at different inclination angles and present the intermediate inclination angle $i=45\degr$, for which the amplitudes for the $|m| = 1$ and $m=0$ modes are comparable.

Figure~\ref{fig:l1_time} shows the results for radial velocity (top left) and the spectral diagnostics for the $l=1, m=-1$ mode (red), $l=1, m=0$ mode (black), and $l=1, m=1$ mode (blue) plotted against time. We fitted sinusoids to the variations (except for FWHM, dLW, and contrast in the case of $m=0$) to highlight the phase relations. The differences between the exact simulated data points and the sinusoids are negligible. 

We first note that all three modes are capable of producing RV variations with similar amplitudes and would thus, in principle, be able to explain the RV variations for NGC 4349 No. 127. While the $|m|=1$ modes are in phase, the $m=0$ mode is phase-shifted. However, we note that an artificial phase shift can be added (and was used to match the retrograde mode with the real observations) to match the $m=0$ mode to the real RVs. 

What cannot be altered, though, are the phase relations between the RV and the spectral diagnostics. First, we observe that for the line shape diagnostics (FWHM, dLW, contrast) the $m=1$ and $m=-1$ modes are roughly in antiphase with each other. The $m=0$ mode shows much smaller line shape variations that are not perfectly sinusoidal (and therefore presented here by connecting the simulated data points) and, most notably, vary at roughly half the RV period. The behavior can be explained as the azimuthal ($\phi$) component of the axisymmetric $m=0$ mode is in all cases 0. Thus, the interplay between the $\phi$ oscillation component and the rotation vector, which causes the majority of the line shape variations as it influences the flanks of the rotationally broadened line profiles, is not present in the $m=0$ mode. The now solely dominating $\theta$ component of the $m=0$ oscillation is symmetric with respect to the rotation axis and thus leads to much smaller line shape variations.

We also show the phase relations and their directions of correlations in Fig.~\ref{fig:supercorrelation} (left column). The dLW, contrast, and FWHM are plotted in the bottom three panels, respectively. We fitted ellipses to the correlations and present the direction of correlation when appropriate. Figure~\ref{fig:supercorrelation} reveals that both $|m| = 1$ modes create closed-loop correlations, while the $m=0$ mode leads to an ``arc-like'' correlation, clearly not consistent with the real data for NGC 4349 No. 127. 

For the two $|m|=1$ modes, we note that the time dependence of the correlation is reversed (indicated by the arrows), a consequence of the phase shifts of the line shape indicators in Fig.~\ref{fig:l1_time}. While the retrograde $m=1$ mode shows an anti-clockwise correlation for the dLW and FWHM (and therefore clockwise for the contrast, as for NGC 4349 No. 127), the prograde $m=-1$ mode correlates with a reversed time dependence and thus cannot explain the real variations. 

It is furthermore evident from Fig.~\ref{fig:l1_time} (bottom left panel), that the CRX is in all cases roughly in phase with the RV with only slightly different amplitudes. These lead to similar positive correlations between CRX and RV in Fig.~\ref{fig:supercorrelation} (top left panel). The similarity of the CRX variations stems from the fact that the wavelength-dependent limb darkening in interplay with the horizontal components at the limb of the star is dominating the CRX amplitude. The CRX variation is therefore, for all $l=1$ modes considered, in phase with the RV variation. This relation holds true as long as the temperature variation $\delta T_\mathrm{eff}$ is small. The CRX-RV correlation thus seems to be a sensitive indicator for the $l$-mode. 

The slight differences in the CRX amplitudes (and therefore the slopes of the correlation) stem from the interplay between the small temperature variation $\delta T_\mathrm{eff}=2.5\,\mathrm{K}$ and the rotational broadening of the spectral lines. That is, the temperature variation by itself has a small effect on the RVs and the CRX, as it slightly increases or decreases the flux from either side of the rotationally broadened line profile, introducing line asymmetries that are measured as RV shifts. As the contrast (flux ratio) between the slightly hotter and cooler halves of the star is less pronounced at longer wavelengths, this effect is wavelength dependent and thus captured by the CRX. For the $m=-1$ and $m=1$ modes, for instance, the temperature variation is phase-shifted by $180\degr$. Thus, the effect of temperature alone enhances the RV and CRX amplitudes in the $m=1$ case, while it decreases both amplitudes in the $m=-1$ case. The CRX-RV correlation for the $m=0$ mode coincides with the effect given by the limb darkening alone.

Figure~\ref{fig:l1_time} finally presents the BIS variations (bottom right panel). Sinusoidal variations at the period of the RV are predicted for all three $m$-modes, albeit with different amplitudes and phase relations. However, as the variations for NGC 4349 No. 127 cannot be detected given the noise of the data, no strong conclusion can be drawn from the BIS indicator in this case. 
All of the above findings are valid at all inclination angles that produce non-zero RV variations for the individual modes. 

The fact that $m=0$ modes lead only to small line shape variations is somewhat discouraging when considering the possibility of applying the same analysis to other evolved stars, suggested to be false positive planet hosts. If an $m=0$ mode is equally likely, and considering the typical scatter in the real time series of activity indicators, it seems very challenging to detect the line shape variations predicted for this mode. Only the CRX (and BIS) variation and its correlation with the RV are predicted to be similar as for NGC 4349 No. 127 and could provide the most direct hint for non-radial oscillations in the case of an $m=0$ mode. However, if stemming from non-radial oscillations, the amplitude of the CRX variations also depend directly on the amplitude of the oscillation velocity $v_\mathrm{osc}$. Therefore, identifying oscillations based on the CRX at much smaller RV amplitudes could be challenging with the typical data sets. 

In principle, all $m$-modes can be expected to be excited, albeit possibly at different amplitudes and periods in the presence of rotation. For NGC 4349 No. 127, considering only dipole ($l=1$) modes, it seems most straight-forward to explain the data with a single $l=1, m=1$ mode. In reality though, a combination of all three $m$-modes could be the most plausible scenario. Such a combination, however, could also be difficult to detect. 

If all three $m$-modes are excited to the same amplitude, and neglecting at first the change in oscillation periods due to rotation, cancellation effects between the $m=1$ and $m=-1$ modes remove the largest part of the variations present in FWHM, contrast, and dLW (as these are in antiphase for the $m=1$ and $m=-1$ modes), while resulting in a large RV variation. Only the variations in CRX would likely remain above a detectable threshold, a behavior that we confirmed with test simulations. The resulting RV curves of a combination of all $m$-modes can be well approximated as the simple sum of the individual RV curves for each mode shown in Fig.~\ref{fig:l1_time}. Therefore, a combination of all three modes at equal and constant amplitudes leads to a phase shift of the RV curve.

In the presence of rotation, the frequencies $\nu_{l,m}$ for different azimuthal quantum numbers $m$ are split (see, e.g., \citealt{Aerts2021}). This frequency split can be approximated as 
\begin{equation}
    \nu_{l,m} - \nu_{l,0} = \frac{m}{P_{\mathrm{rot}}}\left(1-\frac{1}{l(l+1)}\right)
\end{equation} \citep{Chen2017, Brickhill1975}.
For oscillation periods comparable to the rotation period, this can lead to large period changes. 

 A superposition of sinusoidal signals at different periods and amplitudes can lead to RV curves potentially resembling multi-planetary signals. Moreover, if such signals are insufficiently sampled in context of exoplanet surveys and interpreted as single-planet signals, a change in period, amplitude, or phase of the RV curve might be deduced. This could potentially offer an explanation for the amplitude changes and phase shifts detected in the cases of $\gamma$ Dra \citep{Hatzes2018} and Aldebaran \citep{Reichert2019}, although more thorough modeling would be necessary. We also note that a period modulation has recently been reported for some long-period variables showing long secondary periods \citep{Takayama2023}. Of course, in the context of oscillations, it is also plausible to assume that the oscillation periods themselves are not constant.

\subsection{Higher-order l-modes}
While \citet{Saio2015} discuss that dipole oscillatory convective modes have properties that resemble the period-luminosity relations of sequence D variable giant stars, higher-order $l$-modes could also cause false positive planet detections. To understand qualitatively how different quantum numbers $l$ affect the simulation results, we simulated (with the same settings) all possible modes up to $l=4$. 

Motivated by the finding that the $l=1, m=1$ is the only dipole mode consistent with the data for NGC 4349 No. 127, we plot the correlations between the indicators and the RV for the $m=l$ modes in the central column of Fig.~\ref{fig:supercorrelation}. We also show the variations plotted against time in Fig.~\ref{fig:l=m_time}. As in the previous section, we present the results at the inclination angle $i=45\degr$, but note that the results are qualitatively consistent across all inclination angles that produce significant amplitudes in the RVs. We fitted linear correlations for the CRX plotted against RV (top) and elliptical relations for the line shape indicators. All relations follow the simulated data points closely, such that the latter were omitted. Only slight deviations from the linear CRX correlations are present. However, as these are much smaller than the scatter in the real data, we still approximated these with linear relations for visual clarity.

We first note that different RV amplitudes are predicted. While naively one would expect the RV amplitude to decrease for higher-order $l$-modes due to increasing cancellation effects, the behavior is not as straight-forward for non-radial oscillations dominated by the horizontal components. This behavior can be understood as the partial derivatives in Equ. \ref{Eq:vtheta} and \ref{Eq:vphi} introduce a factor $m$ that increases the respective amplitudes as well as (slightly) different normalization factors that were introduced in Equ. \ref{Eq:vr}. The physical horizontal velocity components therefore differ for different modes with the same input velocity $v_\mathrm{osc}$. As $v_\mathrm{osc}$ is a-priori not well constrained, differences in the RV amplitudes can be overcome when attempting to match a mode to the real data. 

Furthermore, sinusoidal variations at the period of the RV are predicted for all indicators. The phase differences between the indicators and the RVs again provide insights to identify the modes. For the line shape indicators FWHM and dLW, the phase shift is positive for the $l=1, m=1$ mode, while it is negative for the higher-degree $m=l$ modes. The behavior of the contrast is reversed. This leads to an inversion of the temporal direction of the elliptical correlations in Fig.~\ref{fig:supercorrelation}. This already disqualifies the higher-degree $m=l$ modes from being consistent with NGC 4349 No. 127. 

However, we find the opposite behavior for the $m=-l$ modes, plotted in the right column of Fig.~\ref{fig:supercorrelation} and Fig.~\ref{fig:l=-m_time}. For $l\geq2$, modes with $m=-l$ produce sinusoidal variations with an anti-clockwise correlation for dLW and FWHM with the RV. The amplitudes of these line shape variations generally depend on the oscillation amplitude $v_\mathrm{osc}$ and the rotation velocity $v_\mathrm{rot}$ and could therefore (to some extent) be adapted in an attempt to match NGC 4349 No. 127. From this finding alone, these higher-order modes would therefore be able to explain the observables for NGC 4349 No. 127. 

Again, we find the CRX to be the most decisive indicator to identify the mode. As presented in the top panels of Fig.~\ref{fig:supercorrelation}, the CRX correlates linearly with the RV for all modes. However, the slope of the correlation depends critically on the mode. While the dipole $l=1$ (red) mode has a positive slope very similar to the real data, the $l=2$ (orange) mode has a decreased but still positive slope. Increasing the quantum number $l$ decreases the slope further, making it progressively negative for $l=3$ (black) and $l=4$ (blue). This behavior is nearly identical for both $m=l$ and $m=-l$ modes, and is roughly consistent for all $m$-modes for the same order $l$. The same effect is also evident in the phase shifts between the different CRX time series and the RVs in Fig.~\ref{fig:l=m_time}. 

The behavior can be understood due to the geometry of the projected and combined oscillation velocities in interplay with the wavelength-dependent limb darkening. The modes are in all cases dominated by the horizontal components (due to the high $K$ factor). For the low-order $m=l=1$ mode, maximum RV was reached when the projected $\phi$ component of the oscillation was fully positive. The wavelength-dependent limb darkening then decreases the overall maximum RV to a larger extent in the blue vs. the red part of the spectrum. This leads to a chromatic RV, captured by a maximum positive CRX at time of maximum positive RV, and therefore the observed positive chromaticity. 

For modes with $l=|m|=3$ and $l=|m|=4$ the dominating $\phi$ component facing the observer is never fully positive. Maximum RV is then reached when the central part of the disk is receding from the observer, while at the same time thin strips at the limb of the star have negative RV values. As these thin strips of cells are most affected by the limb darkening, its wavelength-dependent reduction of RV now has the inverse effect. That is, it diminishes the negative parts and thus increases the overall RV. As this effect is more prominent in the blue vs the red, the phase of maximum CRX is now phase-shifted by half a period with respect to the phase of maximum RV. As a consequence, a negative CRX-RV correlation is observed. 

The $l=|m|=2$ mode is  intermediate. While the $\phi$ component can still entirely be positive, the $\theta$ component already presents a similar effect as detailed for the $l>2$ modes. As the $\theta$ component is generally smaller in summed projected velocities, this acts only to decrease the CRX-RV slope. Of course, the exact CRX-RV slopes depend on the choice of limb-darkening parameters, but the overall trend can be expected to be insensitive to the exact set of parameters. 

To further study this behavior, we simulated even higher $l$-modes, observing that the slopes start to alternate between uneven (positive slope) and even (negative slope) $l$-modes, as can be expected from geometrical considerations. However, due to the ever increasing cancellation effects, simulated RV and indicator amplitudes quickly drop and would thus require much larger values of $v_\mathrm{osc}$ to be detectable.  

Finally, we note that these considerations are only valid for small temperature variations (here $\delta T_\mathrm{eff} = 2.5\,\mathrm{K}$), for which the limb darkening dominates the chromatic behavior. Increased temperature variations change the CRX-RV slope and thus make mode identification based on CRX measurements ambiguous. 

As the overall behavior of the $l=2, m=-2$ mode resembles the retrograde, dipole mode discussed in Sect.~ \ref{sec:results}, we tested whether we could reproduce the data for NGC 4349 No. 127 with this mode. While the CRX-RV correlation can be reproduced with a slightly increased temperature variation $\delta T_\mathrm{eff} \sim 4\,\mathrm{K}$, the amplitudes of the line shape indicators, especially the FWHM and contrast of the CCF, were somewhat smaller than presented for the $l=1, m=1$ simulation for all sets of parameters considered. However, given the scatter of the real data, they provide a fit nearly as good as presented in Sect.~\ref{sec:results}. Due to the increased rotation velocity ($v_{\mathrm{rot}} = 4.5\ \mathrm{\frac{km}{s}}$), necessary to increase the amplitude of the FWHM variations, and the altered oscillation geometry, the amplitude of the BIS variation is increased by a factor $\sim9.6$ compared to Fig.~\ref{fig:timeseries}, yielding a sinusoidal variation slightly larger but still comparable to the scatter in the BIS data. It is unclear whether this signal would have been picked up in the HARPS data set. We therefore cannot exclude the possibility of an $l=2, m=-2$ mode as the cause of the variations.

In general, we find that most modes with $m \neq 0$ for $l\leq4$, produce similar sinusoidal variations of the RVs and indicators which lead to elliptical correlation plots with reversed directions for the two opposite $m$-modes. One exception are $|m|=1$ for $l\geq3$ at specific inclination angles that resemble the arc-like correlations for $m=0$. However, examining this behavior more closely is beyond the scope of this work. The CRX-RV slopes are always very similar for all $m$-modes for one quantum number $l$. The $m = 0$ modes always show a similar behavior to the $l=1, m=0$ mode presented in the previous section and can thus always be refuted for NGC 4349 No. 127.

\section{Discussion}
\label{sec:discussion}
The aim of this study was to test whether non-radial oscillations are capable of explaining the variations of the RVs and the activity indicators of the cluster giant NGC 4349 No. 127. With the data and model presented in Sect.~\ref{sec:results}, we can confirm that the variations of the RVs and indicators are indeed of intrinsic origin and are consistent with a model of a retrograde, dipole oscillation mode. Here, we put these findings into the context of previously reported false positive exoplanets and discuss whether the non-radial oscillations could be oscillatory convective modes. We further discuss that magnetic activity, as suggested by \citet{DelgadoMena2018}, is incompatible with the observations. 

\subsection{Oscillatory convective modes}
\label{sec:discussion:ocm}
\begin{figure*}
    \sidecaption
    \includegraphics{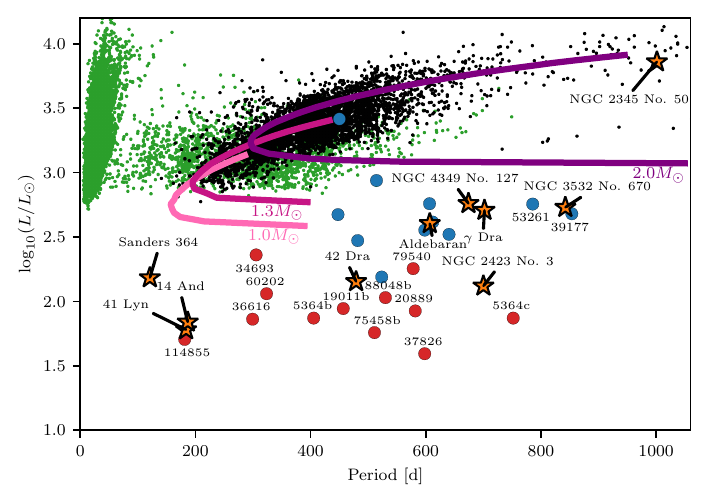}
    \caption{Period-luminosity plot adapted from \citet{Reichert2019} showing the location of giant stars with RV variations of unknown intrinsic origins (orange star markers). The green dots are photometrically variable giant stars identified by the OGLE survey, while the black dots are the subset of these showing long secondary periods. Overplotted in shades of purple are models of oscillatory convective modes (mixing length parameter of 1.2), as presented by \citet{Saio2015}. We also overplot confirmed planets orbiting giant stars from the Lick RV survey (see, e.g., \citealt{Wolthoff2022}) with their respective \textsc{Hipparcos} identifiers in red and (mostly unpublished) planet candidates from the same RV survey in blue. NGC 4349 No. 127 lies in close vicinity to some stars with established intrinsic RV variations, suggesting a common origin for the phenomenon. This impression is further enhanced by the clear separation of the regions in which the confirmed planets and planet candidates fall. }
    \label{fig:pl}
\end{figure*}

In Fig.~\ref{fig:pl}, we adapt Fig.~8 of \citet{Reichert2019}, showing the location of NGC 4349 No. 127 along with the locations of other giant stars (orange star markers) with established intrinsic RV variations (some of which were formerly thought to host planets or brown dwarfs) in a period-luminosity plot. We do not include $\epsilon$ Cyg, as the RV variations of the star are likely caused by the heartbeat phenomenon \citep{Heeren2021}. Luminosities and periods were taken from the respective publications that refute the planets. For Sanders 364, the luminosity was taken from the {\it Gaia} DR3 determination of astrophysical parameters \citep{Creevey2023}. We note that \citet{Zhou2023} find RV periodicity at several periods (224 d, 281 d, 333 d, 530 d, 777 d) longer than the published planet period at 121 d by \citet{Brucalassi2014a, Brucalassi2017}, such that the position of the star could move significantly parallel to the x-axis of Fig.~\ref{fig:pl}.

It is evident that NGC 4349 No. 127 lies in close vicinity to the refuted planet hosts $\gamma$ Dra \citep{Hatzes2018} and Aldebaran \citep{Reichert2019}. We also show the locations of variable giant stars in the Large Magellanic Cloud from the OGLE survey \citep{Soszynski2009} in green and stars which show long secondary periods in black (see \citealt{Reichert2019} for details). We further plot the models taken from \citet{Saio2015} for a mixing length parameter of $\alpha=1.2$. 

Overplotted in red are known planets from the Lick survey (see a recent overview in \citealt{Wolthoff2022}), along with their \textsc{Hipparcos} identifiers, and planet candidates identified in the same survey in blue. The latter group has (semi-)periodic RV signatures, but either high RV jitter or otherwise complicated RV curves, which are challenging to model with Keplerian orbits. The two systems \object{HIP 53261} and \object{HIP 39117} were classified as planet candidates by \citet{TalaPinto2020} and are also listed with their \textsc{Hipparcos} identifiers. 

There appears to be a perceptible distinction between the groups of confirmed planets on the one hand and planet candidates and false positives on the other. The latter mostly accumulate close to the region with periods between $500\,\mathrm{d}$ and $1000\,\mathrm{d}$ and luminosities $2.3 \lesssim \log_{10}(L/L_\sun) \lesssim 3.2$. We find two refuted planetary systems, 42 Dra and NGC 2423 No. 3, to be quite close to established planet hosts. We also find a group of three false positive planet hosts (Sanders 364, 41 Lyn, 14 And) to be located at much lower luminosities and much shorter periods.  As \citet{Reichert2019}, we note that increased luminosity is also linked to increased levels of RV jitter, which makes the confirmation of exoplanets more challenging. 

However, increased RV jitter alone cannot explain the individual cases of most refuted planet systems, which show large-amplitude RV variations, such as the star discussed in this work. As \citet{Hatzes2018} and \citet{Reichert2019} argue, if one were to extend the low-mass models of oscillatory convective modes toward longer periods, they would presumably cross the region covered by Aldebaran, NGC 4349 No. 127, $\gamma$ Dra, and NGC 3532 No. 670. Another cluster giant, NGC 2345 No. 50, appears to be close to the high-luminosity end of the Saio models. 42 Dra, NGC 2423 No. 3, Sanders 364, 41 Lyn, and 14 And seem to have  luminosities that are too low to be in good agreement with the Saio models. As the last three also exhibit shorter periods, one can suspect that another mechanism could be present in these stars, such as the rotational modulation of stellar spots as suggested by \citet{Teng2023b}.

We also note that many of the stars with longer-period RV variations, including NGC 4349 No. 127 ($M=3.01\pm0.24\,M_\sun$), are significantly more massive than predicted by the Saio models. These include $\gamma$ Dra ($M=2.14\pm0.16\,M_\sun$; \citealt{Hatzes2018}), NGC 3532 No. 670 ($M=3.05\pm0.23\,M_\sun$), NGC 2423 No. 3 ($M=2.03\pm0.14\,M_\sun$), and NGC 2345 No. 50 ($M=5.84\pm0.61\,M_\sun$)\citep{Tsantaki2023}. All of these would need to have higher luminosities to be consistent with the Saio models. Only Aldebaran ($M=0.91^{+0.04}_{
-0.02}\,M_\sun$; \citealt{Stock2018}) and 42 Dra ($M=0.98\pm0.05\,M_\sun$; \citealt{Dollinger2009}) have lower masses. Therefore, only Aldebaran is in good agreement with the low-mass models of \citet{Saio2015}. 

For NGC 4349 No. 127, as detailed in Sect.~\ref{sec:observations}, a lower mass $M=2.20^{+0.38}_{-0.20}\,M_\sun$ is also possible, which is however also linked to a lower luminosity $\log_{10}(L/L_\sun) \sim 2.3$. We also note that the models could be altered significantly by a more sophisticated treatment of convection \citep{Saio2015}.

Overall, we find that the properties (mode, amplitude of temperature variation) of the presented oscillation model seem to be consistent with the predictions for oscillatory convective modes. We nevertheless are cautious to conclude that the Saio models provide a good match given the stellar properties and period of the variation. More detailed modeling of their properties would be very valuable. 

As discussed by several authors (e.g., \citealt{Reichert2019, Dollinger2021, DelgadoMena2023}), a suspicious accumulation of false positive planet detections in the vicinity of NGC 4349 No. 127 is evident. As the non-radial oscillation model convincingly explains the variations of the star, a generalization to other stars with similar parameters seems plausible. 

On the other hand, an alternative explanation for the long secondary periods of stars on sequence D is binarity. In this scenario, a brown dwarf or low-mass stellar companion enveloped by a comet-like dust cloud causes the photometric variations by obscuring the stellar disk \citep{Soszynski2007, Soszynski2014}, a proposal supported by the detection of secondary eclipses in mid-infrared photometry for a subsample of variable stars showing long secondary periods \citep{Soszynski2021}. 

While a brown dwarf is indeed consistent with the RVs, this scenario nevertheless seems implausible for NGC 4349 No. 127, as it would not be clear how the line shape and CRX variations could be explained. If we, for the sake of the argument, assume that the brown dwarf exists, is engulfed in a dusty cloud, and is transiting the star, then transit durations are on the order of $T_\mathrm{transit}\sim 22\,\mathrm{d}$, even if one were to assume a radius of $10\,R_\sun$ for the dust cloud. In principle, the Rossiter-McLaughlin effect can explain the line shape variations during the transit itself. But since we observe a near sinusoidal variation, for instance, for the FWHM, the variations out of transit would remain unaccounted for. It would be even more problematic in the cases of Aldebaran or $\gamma$ Dra, which show RV amplitude changes or phase shifts that are hard to explain with an orbital companion. This scenario can thus not explain the observables. 

It is, of course, possible that the variability on sequence D, mostly detected via photometry, is simply not related to the intrinsic RV variations of the discussed giants. Moreover, it is possible that even within the small group of these false positive planet hosts, different intrinsic mechanism are at work. For NGC 4349 No. 127, non-radial oscillations provide the most convincing explanation for all observables to date. Whether these are linked to oscillatory convective modes or not remains uncertain. 

\subsection{Magnetic activity}
\citet{DelgadoMena2018} propose the rotational modulation of magnetic spots as the origin of the RV variations. However, the available photometry and the newly presented activity indicators lead us to refute the spot hypothesis. 

If magnetic surface spots cause the RV periodicity, significant photometric variations are to be expected. Following \citet{Hatzes2002}, we can roughly estimate the required spot filling factor to produce RV variations with amplitude $A_\mathrm{RV}=235.6\,\mathrm{\frac{m}{s}}$ (see Fig.~\ref{fig:rv}) as $f=7.3\%$, using $v\sin i=4.81\,\mathrm{\frac{km}{s}}$ \citep{Tsantaki2023}. \citet{Hatzes2002} used $T_\mathrm{eff}=5800\,\mathrm{K}$ and $T_\mathrm{spot}=4600\,\mathrm{K}$ for the effective temperatures of the stellar photosphere and the star spot, respectively. Downscaling the spot temperature to yield the same flux ratio as considered by \citet{Hatzes2002}, but at effective temperature $T_\mathrm{eff}=4417\,\mathrm{K}$ \citep{Tsantaki2023}, yields a temperature difference $\Delta T_\mathrm{spot}=914\,\mathrm{K}$ between photosphere and spot. Assuming black body radiation, we calculate the flux in a Bessel V-band filter for a photosphere with a star spot of filling factor $f$ to be reduced by $5.7\%$ compared to the flux of a photosphere with no spot present. This corresponds to a magnitude difference of $\Delta V \sim 0.06\, \mathrm{mag}$. 

We searched for the largest amplitude consistent with the ASAS-3 V-band photometry by fitting sinusoids with fixed periods in a box of size $100\,\mathrm{d}$ around the best RV period $P=674.0\,\mathrm{d}$. The largest amplitude present in the data is $A_\mathrm{V}=0.006\pm0.003\,\mathrm{mag}$, an order of magnitude lower than predicted by our rough estimate. We further show in Appendix~\ref{sec:app:photometry} that the apparent periodicity at roughly half the RV period is caused by the window function of the ASAS-3 observations. We therefore conclude that there are no photometric variations at periods related to the RV period in the ASAS-3 data set that are consistent with the prediction.

Another strong argument against the RV variations being caused by star spots is the positive correlation between CRX and RV in Fig.~\ref{fig:corr} (top left panel). If spots were to cause the RV variation, one would expect the CRX to be anticorrelated with the RV (negative chromaticity), as the contrast (flux ratio) between the cool spot and the stellar photosphere decreases toward longer wavelengths \citep{Reiners2010, Barnes2011, Tal-Or2018, Baroch2020, Lafarga2021}. The same relation holds for hot active regions \citep{Reiners2013}. Thus, the positive chromaticity observed for NGC 4349 No. 127 is challenging to explain with a spot model, while it is a natural consequence of the limb darkening in case of dipole non-radial oscillations.

We note that \citet{DelgadoMena2023} recently discussed other magnetic structures that locally reduce convection and lead to RV variations without photometric variations. However, it is not clear if such structures could explain any of the correlations present for NGC 4349 No. 127. We therefore find the oscillation hypothesis more convincing.

\section{Summary}
\label{sec:summary}
We presented a reanalysis of archival HARPS spectra of the cluster giant NGC 4349 No. 127, which was originally thought to host a brown dwarf orbital companion \citep{Lovis2007}. \citet{DelgadoMena2018, DelgadoMena2023}, however, refute the brown dwarf's existence based on variations of the FWHM of the CCF and the $\mathrm{H}\alpha$ index. This makes NGC 4349 No. 127 another example of a group of giant stars with RV periodicities that mimic orbital companions, but are of unknown intrinsic origin. 

We first presented additional activity indicators available through the use of the SERVAL and RACCOON reduction pipelines, revealing that, along with the FWHM,   the contrast of the CCF, the dLW, and the CRX also have significant periodicity at the proposed orbital period. We therefore affirm the findings of \citet{DelgadoMena2018, DelgadoMena2023} and unequivocally refute the orbital companion. Furthermore, we find a significant, positive correlation between the CRX and the RVs, as well as closed-loop correlations between the RVs and dLW, FWHM, and contrast, respectively, a result of sinusoidal variations with phase shifts close to $\pi/2$.

We presented the simulation software \texttt{pyoscillot}, capable of simulating synthetic HARPS spectra, along with their respective RVs and activity indicators, based on a model of non-radial oscillations in the photosphere of a star. Such non-radial oscillations were previously proposed to cause the intrinsic RV variations that mimic orbital companions. 

We showed that an isolated, low-amplitude, retrograde, dipole ($l=1, m=1$) oscillation reproduces the periodicities and amplitudes of the RVs and all activity indicators, along with their respective phase relations and thus the Lissajous figures of their correlations. The oscillation model also predicts low-amplitude photometric variations at the orbital period, which are well below the detection threshold of the available ASAS-3 data, and thus cannot be ruled out.

We also showed that other dipole ($l=1$) modes, while producing similar RV variations, are inconsistent with the activity indicators measured for NGC 4349 No. 127 and present the expected variations for all modes. We showed that the CRX is the only indicator yielding variations for all azimuthal orders $m$ allowed for dipole ($l=1$) modes, while the line shape diagnostics are mostly inconspicuous for the $l=1, m=0$ mode. In the presence of scatter in the CRX data, such modes and combinations of different $l=1$ modes could therefore be difficult to detect. 

Moreover, we explored higher-order modes, showing that modes with $m=l$ and $m=-l$ show similar behavior to the $l=1, m=1$ mode.  The $l=2, m=-2$ mode cannot entirely be ruled out as the  cause of the observed variations. In general, the CRX provides the most direct approach to differentiate between modes, but it is also affected by the scale of the temperature variations employed, and  thus cannot be used to rule out modes unambiguously. 

We finally showed that the stellar parameters of NGC 4349 No. 127 are broadly consistent with the models of oscillatory convective modes \citep{Saio2015} put forth to cause non-radial oscillations of bright giant stars. We showed that reliable planet detections, on the one hand, and more uncertain planet systems and false positives, on the other, are separated in a period-luminosity plot. This suggests that the same intrinsic phenomenon might be present in the latter group. We note, however, that the masses of most giant stars with intrinsic RV signatures are higher than predicted by the models of \citet{Saio2015}.
We further give arguments that the correlations observed for NGC 4349 No. 127 cannot be explained by magnetic spots. 

We conclude that the RV variations of NGC 4349 No. 127 are likely caused by a retrograde, dipole ($l=1, m=1$) non-radial oscillation. We also provide, for the first time,  a testable model capable of explaining the intrinsic RV variations discussed in the literature. 

The case of NGC 4349 No. 127 shows the importance of acquiring a set of reliable and precise activity indicators targeting the shape of spectral lines and the wavelength dependence of the RV. However, as $m=0$ modes do not show large variations of the shape of spectral lines, and would only be detected by the CRX, which often suffers from significant scatter, long-term RV monitoring remains crucial in order  to detect any changes in the amplitude, period, or phase of the RV variation. 

Applying the same analysis to other evolved stars with similar luminosities and RV periods could offer valuable insights to constrain their (potential) intrinsic nature. This analysis may thus contribute to verifying the reliability of the planet population orbiting evolved stars. 

\begin{acknowledgements}
We thank the anonymous referee for their helpful comments that improved the quality of this work. We further thank M. Zechmeister for fruitful discussions regarding the SERVAL reduction results. 
      This work was supported by the Deutsche Forschungsgemeinschaft (DFG) within the Priority Program SPP 1992 “The Diversity of Exoplanets” (RE 2694/7-1). Based on data obtained from the ESO Science Archive Facility with DOI: \url{https://doi.eso.org/10.18727/archive/33}. This research has made use of the VizieR catalog access tool and the SIMBAD database, operated at CDS, Strasbourg, France. This work has made use of data from the European Space Agency (ESA) mission
{\it Gaia} (\url{https://www.cosmos.esa.int/gaia}), processed by the {\it Gaia}
Data Processing and Analysis Consortium (DPAC,
\url{https://www.cosmos.esa.int/web/gaia/dpac/consortium}). Funding for the DPAC
has been provided by national institutions, in particular the institutions
participating in the {\it Gaia} Multilateral Agreement. This research made use of Astropy,\footnote{\url{http://www.astropy.org}} a community-developed core Python package for Astronomy \citep{AstropyCollaboration2013, AstropyCollaboration2018}. We further acknowledge use of the Python packages NumPy \citep{Harris2020}, SciPy \citep{Virtanen2020}, Matplotlib \citep{Hunter2007}, pandas \citep{pandas2020, McKinney2010}, PyAstronomy\footnote{\url{https://github.com/sczesla/PyAstronomy}} \citep{Czesla2019}, lsq-ellipse \citep{Hammel2020}, the RV fitting tool Exo-Striker \citep{Trifonov2019a}, and the color map ``vikO'' \citep{Crameri2023}. 
\end{acknowledgements}

\bibliographystyle{aa} 
\bibliography{Literature}

\begin{appendix}
\FloatBarrier
\section{Linear correlation between \texorpdfstring{H$\alpha$}{H alpha} and RV}
\begin{figure}[h]
    \centering
    \includegraphics{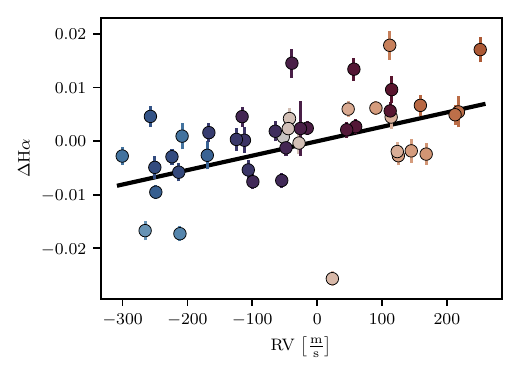}
    \caption{Correlation between H$\alpha$ indicator (absolute difference from the mean) and RV for the 44 spectra acquired prior to the HARPS fiber change. The same color-coding as in Fig.~\ref{fig:corr} applies. A linear relation (black line) was fitted to the data. An \textit{F}-test against a constant model confirms the correlation to be significant with a p-value $p(\textit{F}\textrm{-test})=0.2\%$ and Pearson's r coefficient $r=0.48$.}
    \label{fig:halpha}
\end{figure}

\section{Discussion of the ASAS-3 photometry}
\label{sec:app:photometry}
\begin{figure}[h]
    \centering
    \includegraphics{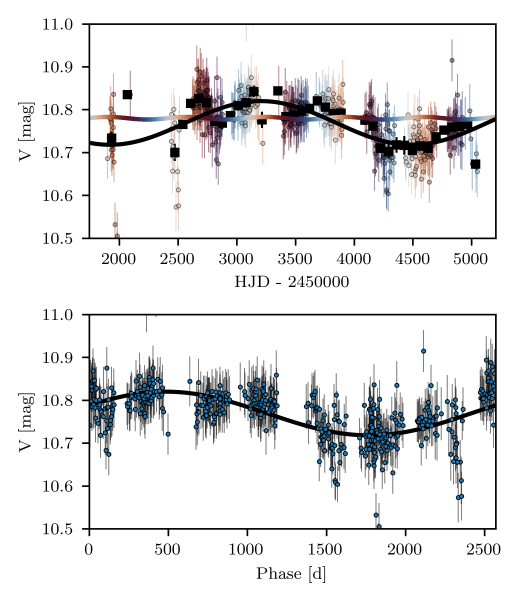}
    \caption{Observed and simulated photometry for NGC 4349 No. 127. Top: V magnitudes extracted from ASAS-3 plotted against time (colored dots). The color-coding is the same as in Fig.~\ref{fig:corr} and refers to the RV period. We also plot binned means (black rectangles) as well as the photometric variability predicted by the $l=1, m=1$ oscillation model (continuous multi-colored line). We further overplot a sinusoidal fit with period $2571\,\mathrm{d}$ in black. Bottom: ASAS-3 V magnitudes phase-folded to the $2571\,\mathrm{d}$ period.}
    \label{fig:phot}
\end{figure}

\begin{figure}[h]
    \centering
    \includegraphics{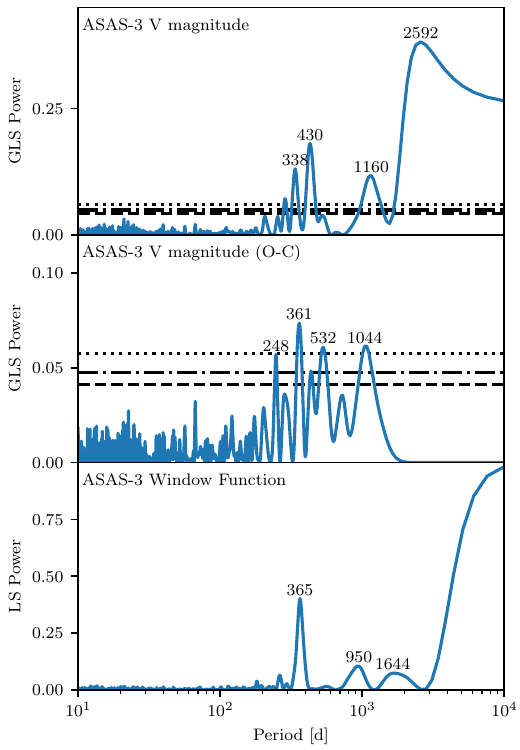}
    \caption{GLS periodogram analysis of the ASAS-3 V-band photometry. Top: GLS periodogram of the original photometry. The FAP levels of $5\%$ (dashed), $1\%$ (dash-dotted), and $0.1\%$ (dotted) were computed via bootstrapping with 10\,000 reshuffles and are plotted as horizontal lines. The periods of the four most prominent peaks are given in units of days. 
    Middle: GLS periodogram of the residual photometry after removing the $2571\,\mathrm{d}$ fit. 
    Bottom: LS periodogram of the window function. }
    \label{fig:phot_gls}
\end{figure}
In Fig.~\ref{fig:phot} (top), we plot the available ASAS-3 data against time, color-coding the individual data points with respect to the RV period. For enhanced visibility, we further show binned data points as black rectangles. The photometric variation predicted by the $l=1, m=1$ oscillation model discussed in Sect.~\ref{sec:results} is overplotted as the colored line and is much smaller than the scatter present in the ASAS-3 data. It is therefore reasonable to assume that the photometric variations would not have been detected in the ASAS-3 data set. 

In Fig.~\ref{fig:phot_gls} (top), we plot the GLS periodogram of the ASAS-3 data, labeling the four most significant peaks in units of days. \citet{DelgadoMena2018} discuss the same data set and argue that the peak at $338\,\mathrm{d}$ is close to the first harmonic ($P/2$) of the RV period or could be linked to the rotation period of the star. However, we find that the most significant peak at $2592\,\mathrm{d}$ provides a much more convincing fit to the ASAS-3 data set, which we plot in black in the top panel of Fig.~\ref{fig:phot}. The sinusoidal fit has an amplitude of 0.05 mag and a period of $2571\,\mathrm{d}$. The discrepancy to the $2592\,\mathrm{d}$ period found in the GLS periodogram stems from the sampling of the periodogram. We show a phase-folded plot of the ASAS-3 data in the bottom panel of Fig.~\ref{fig:phot}.

In the middle panel of Fig.~\ref{fig:phot_gls}, we show the GLS periodogram of the residuals after removing the fit with period $2571\,\mathrm{d}$. The bottom panel further shows the periodogram of the window function computed as a Lomb-Scargle (LS) periodogram, as suggested by \citet{VanderPlas2018}. The formerly second strongest peak at $430\,\mathrm{d}$ coincides with the alias of the $2592\,\mathrm{d}$ signal and the strong yearly period present in the window function. It is therefore much reduced in significance by removing the long-period signal. 

The interesting $338\,\mathrm{d}$ period also changes significantly by removing the fit with period $2571\,\mathrm{d}$. It is reduced in significance and shifted to $361\,\mathrm{d}$, coinciding quite closely with the yearly period. The two neighboring peaks at $248\,\mathrm{d}$ and $532\,\mathrm{d}$ are furthermore close to the alias period of the yearly period and the $950\,\mathrm{d}$ period present in the window function. We therefore argue that the $338\,\mathrm{d}$ signal discussed by \citet{DelgadoMena2018} is likely not real and instead an artifact of the window function, as well as its neighboring peaks. 

The origin of the fourth peak at $1160\,\mathrm{d}$ present in the original GLS periodogram is less certain. While it is close to the first harmonic of the long-period signal, it is not entirely suppressed by removing the $2571\,\mathrm{d}$ fit. It is, however, shifted closer to the $950\,\mathrm{d}$ peak present in the window function and could therefore potentially be explained as another windowing artifact. 

For the $2592\,\mathrm{d}$ peak, there is no obvious peak present in the window function. It is also unlikely to be caused by the typical rise of the periodogram power of the window function at periods longer than the observing window. We therefore argue that the $2592\,\mathrm{d}$ signal is the only signal that cannot be explained by the window function. Whether this signal is of instrumental or stellar origin is uncertain. If it is real, it seems unlikely to be related to the non-radial oscillation discussed in Sect.~\ref{sec:results}. 

It could, in principle, be related to the rotation period of the star. However, given the radius of the star, this would lead to a rotation velocity of $v_{\mathrm{rot}} \sim 0.75\,\mathrm{km\,s^{-1}}$, considerably lower than most estimates in the literature (see Table~\ref{tab:params}), but consistent with the estimate by \citet{Carlberg2016}. We note that the presented simulation of a $l=1, m=1$ oscillation mode also requires a somewhat lower rotation velocity to reproduce the variations of the line shape indicators, but caution to use this value to determine the actual rotation velocity of the star. 

\clearpage
\onecolumn
\section{Oscillation modes with \texorpdfstring{$m=l$}{m=l}}
\begin{figure*}[h]
\sidecaption
    \includegraphics{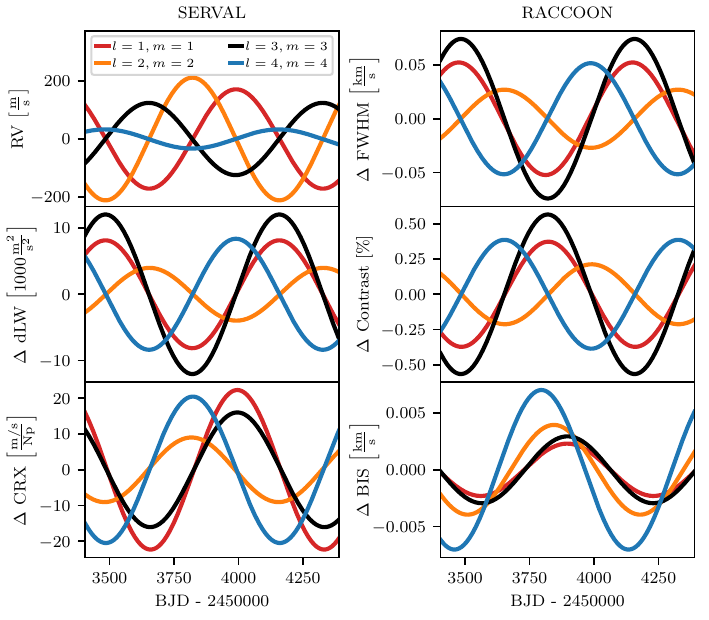}
     \caption{RVs and activity indicators for oscillation modes with $m=l$ plotted against time at inclination angle $i=45\degr$. Sinusoids are predicted for all observables and were fitted to the time series. Different amplitudes and phase relations are predicted and can be useful for mode identification. The dLW variations were rescaled with a common factor of $0.1$.}
    \label{fig:l=m_time}
\end{figure*}
\section{Oscillation modes with \texorpdfstring{$m=-l$}{m=-l}}
\begin{figure*}[h]
    \sidecaption
    \includegraphics{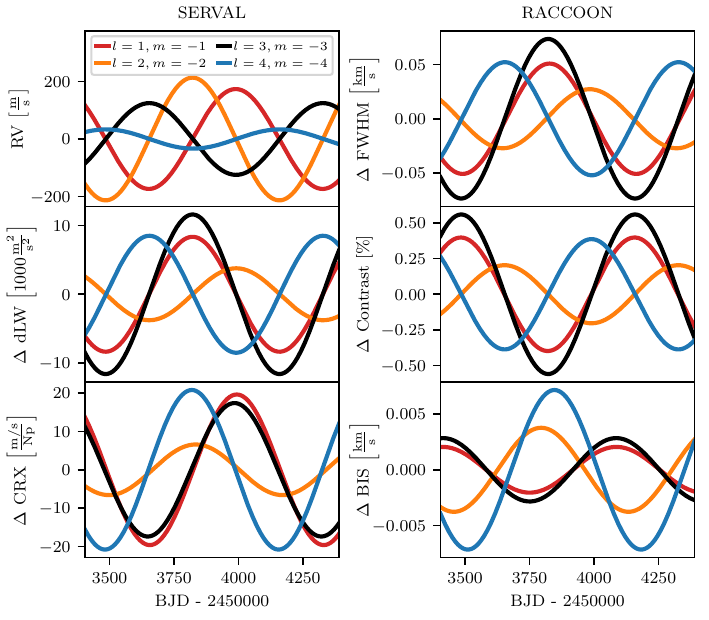}
    \caption{RVs and activity indicators for oscillation modes with $m=-l$ plotted against time at inclination angle $i=45\degr$. Sinusoids are predicted for all observables and were fitted to the time series. Different amplitudes and phase relations are predicted and can be useful for mode identification. The dLW variations were rescaled with a common factor of $0.1$.}
    \label{fig:l=-m_time}
\end{figure*}
\FloatBarrier
\clearpage
\section{HARPS data set}
\begin{table*}[h]
\caption{SERVAL and RACCOON reduction results.}
\label{tab:data}
\resizebox{\textwidth}{!}{
\begin{tabular}{lllllllll}
\hline
 \phantom{(}BJD & \phantom{(}RV  & \phantom{(}RV & \phantom{(}CRX & \phantom{(}dLW & \phantom{(}H$\alpha$ & \phantom{(}FWHM & \phantom{(}Contrast & \phantom{(}BIS\\
 \phantom{(}SERVAL & \phantom{(}SERVAL & \phantom{(}RACCOON & \phantom{(}SERVAL & \phantom{(}SERVAL & \phantom{(}SERVAL & \phantom{(}RACCOON & \phantom{(}RACCOON & \phantom{(}RACCOON\\
 & \phantom{(}$\frac{\mathrm{m}}{\mathrm{s}}$ & \phantom{(}$\frac{\mathrm{m}}{\mathrm{s}}$ & \phantom{(}$\frac{\mathrm{m}/\mathrm{s}}{\mathrm{Np}}$ & \phantom{(}$1000\frac{\mathrm{m^2}}{\mathrm{s^2}}$ & & \phantom{(}$\frac{\mathrm{km}}{\mathrm{s}}$ & \phantom{(}$\%$ & \phantom{(}$\frac{\mathrm{km}}{\mathrm{s}}$\\
\hline
$\phantom{(}2453449.783796\phantom{)}$ & $\phantom{(}\phantom{\mathopen{}-}127.1\pm1.9\phantom{)}$ & $\phantom{(}\mathopen{}-11528.8\pm\phantom{0}6.8\phantom{)}$ & $\phantom{(}\phantom{0}\mathopen{}-7.5\pm14.8\phantom{)}$ & $\phantom{(}\phantom{\mathopen{}-}19.9\pm1.7\phantom{)}$ & $\phantom{(}0.4091\pm0.0017\phantom{)}$ & $\phantom{(}9.205\pm0.019\phantom{)}$ & $\phantom{(}49.777\pm0.077\phantom{)}$ & $\phantom{(}0.076\pm0.016\phantom{)}$\\
$\phantom{(}2453460.836163\phantom{)}$ & $\phantom{(}\phantom{0}\phantom{\mathopen{}-}71.1\pm2.8\phantom{)}$ & $\phantom{(}\mathopen{}-11578.2\pm\phantom{0}8.8\phantom{)}$ & $\phantom{(}\phantom{\mathopen{}-}55.8\pm20.9\phantom{)}$ & $\phantom{(}\phantom{\mathopen{}-}18.6\pm2.6\phantom{)}$ & $\phantom{(}0.4169\pm0.0022\phantom{)}$ & $\phantom{(}9.228\pm0.022\phantom{)}$ & $\phantom{(}48.757\pm0.087\phantom{)}$ & $\phantom{(}0.095\pm0.021\phantom{)}$\\
$\phantom{(}2453469.790637\phantom{)}$ & $\phantom{(}\phantom{0}\phantom{\mathopen{}-}60.0\pm2.4\phantom{)}$ & $\phantom{(}\mathopen{}-11598.1\pm\phantom{0}6.0\phantom{)}$ & $\phantom{(}\phantom{\mathopen{}-}15.4\pm18.1\phantom{)}$ & $\phantom{(}\phantom{0}\phantom{\mathopen{}-}5.0\pm1.8\phantom{)}$ & $\phantom{(}0.4056\pm0.0015\phantom{)}$ & $\phantom{(}9.157\pm0.016\phantom{)}$ & $\phantom{(}49.700\pm0.065\phantom{)}$ & $\phantom{(}0.059\pm0.014\phantom{)}$\\
$\phantom{(}2453499.578838\phantom{)}$ & $\phantom{(}\phantom{0}\mathopen{}-24.5\pm3.4\phantom{)}$ & $\phantom{(}\mathopen{}-11688.5\pm10.8\phantom{)}$ & $\phantom{(}\phantom{0}\phantom{\mathopen{}-}8.6\pm26.9\phantom{)}$ & $\phantom{(}\phantom{\mathopen{}-}14.6\pm4.1\phantom{)}$ & $\phantom{(}0.4180\pm0.0027\phantom{)}$ & $\phantom{(}9.168\pm0.020\phantom{)}$ & $\phantom{(}48.874\pm0.078\phantom{)}$ & $\phantom{(}0.043\pm0.025\phantom{)}$\\
$\phantom{(}2453500.641879\phantom{)}$ & $\phantom{(}\phantom{0}\mathopen{}-11.3\pm9.1\phantom{)}$ & $\phantom{(}\mathopen{}-11654.9\pm19.4\phantom{)}$ & $\phantom{(}\phantom{\mathopen{}-}36.6\pm71.7\phantom{)}$ & $\phantom{(}\phantom{\mathopen{}-}19.5\pm6.3\phantom{)}$ & $\phantom{(}0.4058\pm0.0051\phantom{)}$ & $\phantom{(}9.215\pm0.020\phantom{)}$ & $\phantom{(}48.047\pm0.076\phantom{)}$ & $\phantom{(}0.066\pm0.044\phantom{)}$\\
$\phantom{(}2453521.584883\phantom{)}$ & $\phantom{(}\mathopen{}-101.3\pm2.7\phantom{)}$ & $\phantom{(}\mathopen{}-11754.8\pm\phantom{0}7.7\phantom{)}$ & $\phantom{(}\mathopen{}-18.4\pm20.4\phantom{)}$ & $\phantom{(}\phantom{\mathopen{}-}10.9\pm2.3\phantom{)}$ & $\phantom{(}0.4080\pm0.0019\phantom{)}$ & $\phantom{(}9.184\pm0.018\phantom{)}$ & $\phantom{(}48.917\pm0.070\phantom{)}$ & $\phantom{(}0.069\pm0.018\phantom{)}$\\
$\phantom{(}2453787.795108\phantom{)}$ & $\phantom{(}\phantom{0}\mathopen{}-37.8\pm2.0\phantom{)}$ & $\phantom{(}\mathopen{}-11698.1\pm\phantom{0}6.2\phantom{)}$ & $\phantom{(}\phantom{0}\mathopen{}-6.9\pm15.5\phantom{)}$ & $\phantom{(}\mathopen{}-20.4\pm1.9\phantom{)}$ & $\phantom{(}0.4042\pm0.0016\phantom{)}$ & $\phantom{(}9.015\pm0.016\phantom{)}$ & $\phantom{(}50.623\pm0.067\phantom{)}$ & $\phantom{(}0.059\pm0.014\phantom{)}$\\
$\phantom{(}2453812.758975\phantom{)}$ & $\phantom{(}\phantom{0}\mathopen{}-28.2\pm2.3\phantom{)}$ & $\phantom{(}\mathopen{}-11687.3\pm\phantom{0}7.3\phantom{)}$ & $\phantom{(}\mathopen{}-28.3\pm17.2\phantom{)}$ & $\phantom{(}\mathopen{}-17.7\pm2.1\phantom{)}$ & $\phantom{(}0.4077\pm0.0019\phantom{)}$ & $\phantom{(}9.027\pm0.016\phantom{)}$ & $\phantom{(}50.444\pm0.066\phantom{)}$ & $\phantom{(}0.052\pm0.017\phantom{)}$\\
$\phantom{(}2453833.703149\phantom{)}$ & $\phantom{(}\phantom{0}\phantom{\mathopen{}-}37.9\pm2.0\phantom{)}$ & $\phantom{(}\mathopen{}-11624.7\pm\phantom{0}5.5\phantom{)}$ & $\phantom{(}\mathopen{}-13.6\pm15.4\phantom{)}$ & $\phantom{(}\mathopen{}-15.0\pm1.6\phantom{)}$ & $\phantom{(}0.3778\pm0.0014\phantom{)}$ & $\phantom{(}9.069\pm0.014\phantom{)}$ & $\phantom{(}50.372\pm0.057\phantom{)}$ & $\phantom{(}0.055\pm0.013\phantom{)}$\\
$\phantom{(}2453862.612919\phantom{)}$ & $\phantom{(}\phantom{0}\phantom{\mathopen{}-}62.3\pm2.1\phantom{)}$ & $\phantom{(}\mathopen{}-11594.8\pm\phantom{0}5.7\phantom{)}$ & $\phantom{(}\phantom{\mathopen{}-}38.2\pm14.9\phantom{)}$ & $\phantom{(}\mathopen{}-14.3\pm1.7\phantom{)}$ & $\phantom{(}0.4094\pm0.0015\phantom{)}$ & $\phantom{(}9.062\pm0.020\phantom{)}$ & $\phantom{(}50.258\pm0.082\phantom{)}$ & $\phantom{(}0.079\pm0.013\phantom{)}$\\
$\phantom{(}2453883.589922\phantom{)}$ & $\phantom{(}\phantom{\mathopen{}-}105.1\pm1.8\phantom{)}$ & $\phantom{(}\mathopen{}-11550.8\pm\phantom{0}5.0\phantom{)}$ & $\phantom{(}\phantom{\mathopen{}-}15.2\pm13.7\phantom{)}$ & $\phantom{(}\mathopen{}-14.1\pm1.4\phantom{)}$ & $\phantom{(}0.4096\pm0.0013\phantom{)}$ & $\phantom{(}9.060\pm0.018\phantom{)}$ & $\phantom{(}50.306\pm0.073\phantom{)}$ & $\phantom{(}0.074\pm0.012\phantom{)}$\\
$\phantom{(}2453922.499565\phantom{)}$ & $\phantom{(}\phantom{\mathopen{}-}126.4\pm4.4\phantom{)}$ & $\phantom{(}\mathopen{}-11532.0\pm10.3\phantom{)}$ & $\phantom{(}\mathopen{}-42.2\pm34.2\phantom{)}$ & $\phantom{(}\phantom{0}\phantom{\mathopen{}-}1.5\pm2.8\phantom{)}$ & $\phantom{(}0.4213\pm0.0027\phantom{)}$ & $\phantom{(}9.158\pm0.019\phantom{)}$ & $\phantom{(}49.554\pm0.075\phantom{)}$ & $\phantom{(}0.076\pm0.024\phantom{)}$\\
$\phantom{(}2453950.476290\phantom{)}$ & $\phantom{(}\phantom{\mathopen{}-}173.6\pm3.3\phantom{)}$ & $\phantom{(}\mathopen{}-11476.0\pm\phantom{0}7.7\phantom{)}$ & $\phantom{(}\phantom{\mathopen{}-}43.5\pm24.5\phantom{)}$ & $\phantom{(}\phantom{0}\mathopen{}-2.2\pm2.2\phantom{)}$ & $\phantom{(}0.4101\pm0.0020\phantom{)}$ & $\phantom{(}9.163\pm0.019\phantom{)}$ & $\phantom{(}49.765\pm0.075\phantom{)}$ & $\phantom{(}0.062\pm0.018\phantom{)}$\\
$\phantom{(}2454117.846012\phantom{)}$ & $\phantom{(}\phantom{\mathopen{}-}129.3\pm4.8\phantom{)}$ & $\phantom{(}\mathopen{}-11514.1\pm\phantom{0}9.7\phantom{)}$ & $\phantom{(}\phantom{\mathopen{}-}13.6\pm37.6\phantom{)}$ & $\phantom{(}\phantom{0}\phantom{\mathopen{}-}4.3\pm2.9\phantom{)}$ & $\phantom{(}0.4131\pm0.0026\phantom{)}$ & $\phantom{(}9.149\pm0.013\phantom{)}$ & $\phantom{(}49.463\pm0.053\phantom{)}$ & $\phantom{(}0.048\pm0.023\phantom{)}$\\
$\phantom{(}2454137.806299\phantom{)}$ & $\phantom{(}\phantom{0}\phantom{\mathopen{}-}73.9\pm2.0\phantom{)}$ & $\phantom{(}\mathopen{}-11582.4\pm\phantom{0}5.7\phantom{)}$ & $\phantom{(}\mathopen{}-22.1\pm14.5\phantom{)}$ & $\phantom{(}\phantom{0}\phantom{\mathopen{}-}6.5\pm1.6\phantom{)}$ & $\phantom{(}0.4061\pm0.0014\phantom{)}$ & $\phantom{(}9.191\pm0.018\phantom{)}$ & $\phantom{(}49.416\pm0.072\phantom{)}$ & $\phantom{(}0.073\pm0.013\phantom{)}$\\
$\phantom{(}2454169.723975\phantom{)}$ & $\phantom{(}\phantom{0}\phantom{0}\mathopen{}-0.8\pm1.5\phantom{)}$ & $\phantom{(}\mathopen{}-11653.4\pm\phantom{0}5.0\phantom{)}$ & $\phantom{(}\phantom{0}\phantom{\mathopen{}-}6.5\pm11.5\phantom{)}$ & $\phantom{(}\phantom{0}\phantom{\mathopen{}-}0.4\pm1.3\phantom{)}$ & $\phantom{(}0.4059\pm0.0013\phantom{)}$ & $\phantom{(}9.178\pm0.017\phantom{)}$ & $\phantom{(}49.463\pm0.068\phantom{)}$ & $\phantom{(}0.061\pm0.012\phantom{)}$\\
$\phantom{(}2454194.779612\phantom{)}$ & $\phantom{(}\phantom{0}\mathopen{}-33.6\pm2.1\phantom{)}$ & $\phantom{(}\mathopen{}-11691.2\pm\phantom{0}6.1\phantom{)}$ & $\phantom{(}\mathopen{}-14.2\pm15.7\phantom{)}$ & $\phantom{(}\phantom{0}\phantom{\mathopen{}-}6.9\pm1.9\phantom{)}$ & $\phantom{(}0.4022\pm0.0016\phantom{)}$ & $\phantom{(}9.207\pm0.016\phantom{)}$ & $\phantom{(}49.253\pm0.063\phantom{)}$ & $\phantom{(}0.053\pm0.014\phantom{)}$\\
$\phantom{(}2454202.725871\phantom{)}$ & $\phantom{(}\phantom{0}\mathopen{}-84.7\pm1.9\phantom{)}$ & $\phantom{(}\mathopen{}-11738.0\pm\phantom{0}5.4\phantom{)}$ & $\phantom{(}\phantom{\mathopen{}-}18.0\pm14.1\phantom{)}$ & $\phantom{(}\phantom{0}\phantom{\mathopen{}-}7.1\pm1.7\phantom{)}$ & $\phantom{(}0.3959\pm0.0014\phantom{)}$ & $\phantom{(}9.184\pm0.017\phantom{)}$ & $\phantom{(}49.591\pm0.067\phantom{)}$ & $\phantom{(}0.059\pm0.013\phantom{)}$\\
$\phantom{(}2454225.659919\phantom{)}$ & $\phantom{(}\phantom{0}\mathopen{}-97.7\pm3.2\phantom{)}$ & $\phantom{(}\mathopen{}-11765.4\pm\phantom{0}9.3\phantom{)}$ & $\phantom{(}\phantom{0}\mathopen{}-8.9\pm24.6\phantom{)}$ & $\phantom{(}\phantom{\mathopen{}-}23.7\pm2.4\phantom{)}$ & $\phantom{(}0.4036\pm0.0024\phantom{)}$ & $\phantom{(}9.258\pm0.016\phantom{)}$ & $\phantom{(}48.711\pm0.062\phantom{)}$ & $\phantom{(}0.053\pm0.022\phantom{)}$\\
$\phantom{(}2454228.669709\phantom{)}$ & $\phantom{(}\phantom{0}\mathopen{}-91.7\pm2.3\phantom{)}$ & $\phantom{(}\mathopen{}-11754.5\pm\phantom{0}7.1\phantom{)}$ & $\phantom{(}\mathopen{}-16.1\pm17.7\phantom{)}$ & $\phantom{(}\phantom{\mathopen{}-}14.7\pm2.3\phantom{)}$ & $\phantom{(}0.3981\pm0.0018\phantom{)}$ & $\phantom{(}9.224\pm0.017\phantom{)}$ & $\phantom{(}49.240\pm0.068\phantom{)}$ & $\phantom{(}0.057\pm0.017\phantom{)}$\\
$\phantom{(}2454233.617177\phantom{)}$ & $\phantom{(}\mathopen{}-110.1\pm2.7\phantom{)}$ & $\phantom{(}\mathopen{}-11763.3\pm\phantom{0}8.1\phantom{)}$ & $\phantom{(}\phantom{0}\phantom{\mathopen{}-}0.6\pm20.8\phantom{)}$ & $\phantom{(}\phantom{\mathopen{}-}16.6\pm1.9\phantom{)}$ & $\phantom{(}0.4038\pm0.0021\phantom{)}$ & $\phantom{(}9.229\pm0.018\phantom{)}$ & $\phantom{(}49.053\pm0.070\phantom{)}$ & $\phantom{(}0.063\pm0.019\phantom{)}$\\
$\phantom{(}2454258.558537\phantom{)}$ & $\phantom{(}\mathopen{}-199.0\pm2.6\phantom{)}$ & $\phantom{(}\mathopen{}-11853.6\pm\phantom{0}6.7\phantom{)}$ & $\phantom{(}\phantom{0}\mathopen{}-7.3\pm19.5\phantom{)}$ & $\phantom{(}\phantom{\mathopen{}-}12.4\pm1.9\phantom{)}$ & $\phantom{(}0.3976\pm0.0017\phantom{)}$ & $\phantom{(}9.209\pm0.024\phantom{)}$ & $\phantom{(}49.347\pm0.095\phantom{)}$ & $\phantom{(}0.101\pm0.016\phantom{)}$\\
$\phantom{(}2454293.529869\phantom{)}$ & $\phantom{(}\mathopen{}-234.3\pm1.9\phantom{)}$ & $\phantom{(}\mathopen{}-11889.5\pm\phantom{0}5.3\phantom{)}$ & $\phantom{(}\mathopen{}-30.9\pm14.3\phantom{)}$ & $\phantom{(}\phantom{0}\mathopen{}-3.4\pm1.5\phantom{)}$ & $\phantom{(}0.3939\pm0.0013\phantom{)}$ & $\phantom{(}9.077\pm0.020\phantom{)}$ & $\phantom{(}49.832\pm0.080\phantom{)}$ & $\phantom{(}0.068\pm0.013\phantom{)}$\\
$\phantom{(}2454299.561525\phantom{)}$ & $\phantom{(}\mathopen{}-155.0\pm3.9\phantom{)}$ & $\phantom{(}\mathopen{}-11807.7\pm10.0\phantom{)}$ & $\phantom{(}\phantom{0}\phantom{\mathopen{}-}9.3\pm30.2\phantom{)}$ & $\phantom{(}\phantom{\mathopen{}-}10.8\pm2.5\phantom{)}$ & $\phantom{(}0.4008\pm0.0026\phantom{)}$ & $\phantom{(}9.125\pm0.017\phantom{)}$ & $\phantom{(}49.289\pm0.068\phantom{)}$ & $\phantom{(}0.050\pm0.023\phantom{)}$\\
$\phantom{(}2454319.474304\phantom{)}$ & $\phantom{(}\mathopen{}-193.8\pm3.8\phantom{)}$ & $\phantom{(}\mathopen{}-11843.4\pm\phantom{0}9.6\phantom{)}$ & $\phantom{(}\mathopen{}-15.5\pm29.7\phantom{)}$ & $\phantom{(}\phantom{0}\phantom{\mathopen{}-}6.7\pm3.2\phantom{)}$ & $\phantom{(}0.4044\pm0.0025\phantom{)}$ & $\phantom{(}9.132\pm0.016\phantom{)}$ & $\phantom{(}49.377\pm0.064\phantom{)}$ & $\phantom{(}0.034\pm0.022\phantom{)}$\\
$(2454323.471811)$ & $(\mathopen{}-205.0\pm4.0)$ & $(\mathopen{}-11866.9\pm\phantom{0}9.3)$ & $(\phantom{\mathopen{}-}28.5\pm32.0)$ & $(\phantom{\mathopen{}-}96.6\pm2.7)$ & $(0.4059\pm0.0024)$ & $(9.114\pm0.018)$ & $(49.211\pm0.071)$ & $(0.063\pm0.022)$\\
$\phantom{(}2454342.475808\phantom{)}$ & $\phantom{(}\mathopen{}-197.1\pm2.4\phantom{)}$ & $\phantom{(}\mathopen{}-11851.7\pm\phantom{0}5.8\phantom{)}$ & $\phantom{(}\mathopen{}-16.2\pm18.0\phantom{)}$ & $\phantom{(}\phantom{0}\phantom{\mathopen{}-}1.9\pm2.0\phantom{)}$ & $\phantom{(}0.3862\pm0.0015\phantom{)}$ & $\phantom{(}9.122\pm0.022\phantom{)}$ & $\phantom{(}49.853\pm0.088\phantom{)}$ & $\phantom{(}0.074\pm0.014\phantom{)}$\\
$(2454349.472032)$ & $(\mathopen{}-182.2\pm2.7)$ & $(\mathopen{}-11839.4\pm\phantom{0}6.5)$ & $(\mathopen{}-13.1\pm21.6)$ & $(\phantom{\mathopen{}-}96.4\pm2.0)$ & $(0.3943\pm0.0016)$ & $(9.130\pm0.018)$ & $(49.691\pm0.072)$ & $(0.070\pm0.015)$\\
$\phantom{(}2454481.828543\phantom{)}$ & $\phantom{(}\phantom{0}\mathopen{}-30.1\pm2.6\phantom{)}$ & $\phantom{(}\mathopen{}-11686.1\pm\phantom{0}6.3\phantom{)}$ & $\phantom{(}\phantom{0}\phantom{\mathopen{}-}3.9\pm19.8\phantom{)}$ & $\phantom{(}\mathopen{}-10.6\pm2.3\phantom{)}$ & $\phantom{(}0.4058\pm0.0017\phantom{)}$ & $\phantom{(}9.033\pm0.021\phantom{)}$ & $\phantom{(}50.538\pm0.087\phantom{)}$ & $\phantom{(}0.067\pm0.015\phantom{)}$\\
$\phantom{(}2454486.799252\phantom{)}$ & $\phantom{(}\phantom{0}\mathopen{}-13.5\pm3.2\phantom{)}$ & $\phantom{(}\mathopen{}-11668.1\pm\phantom{0}7.8\phantom{)}$ & $\phantom{(}\phantom{\mathopen{}-}40.2\pm24.0\phantom{)}$ & $\phantom{(}\phantom{0}\mathopen{}-3.3\pm2.7\phantom{)}$ & $\phantom{(}0.4031\pm0.0021\phantom{)}$ & $\phantom{(}9.046\pm0.019\phantom{)}$ & $\phantom{(}50.143\pm0.079\phantom{)}$ & $\phantom{(}0.087\pm0.018\phantom{)}$\\
$\phantom{(}2454523.803657\phantom{)}$ & $\phantom{(}\phantom{\mathopen{}-}128.6\pm3.2\phantom{)}$ & $\phantom{(}\mathopen{}-11527.4\pm\phantom{0}8.2\phantom{)}$ & $\phantom{(}\phantom{\mathopen{}-}21.7\pm24.6\phantom{)}$ & $\phantom{(}\phantom{0}\mathopen{}-3.8\pm2.8\phantom{)}$ & $\phantom{(}0.4079\pm0.0023\phantom{)}$ & $\phantom{(}9.055\pm0.022\phantom{)}$ & $\phantom{(}50.172\pm0.092\phantom{)}$ & $\phantom{(}0.088\pm0.019\phantom{)}$\\
$\phantom{(}2454528.781976\phantom{)}$ & $\phantom{(}\phantom{\mathopen{}-}138.0\pm2.6\phantom{)}$ & $\phantom{(}\mathopen{}-11513.1\pm\phantom{0}6.6\phantom{)}$ & $\phantom{(}\phantom{\mathopen{}-}48.7\pm18.9\phantom{)}$ & $\phantom{(}\mathopen{}-13.0\pm2.1\phantom{)}$ & $\phantom{(}0.4015\pm0.0018\phantom{)}$ & $\phantom{(}9.039\pm0.022\phantom{)}$ & $\phantom{(}50.619\pm0.092\phantom{)}$ & $\phantom{(}0.095\pm0.016\phantom{)}$\\
$\phantom{(}2454553.728094\phantom{)}$ & $\phantom{(}\phantom{\mathopen{}-}139.7\pm2.6\phantom{)}$ & $\phantom{(}\mathopen{}-11518.1\pm\phantom{0}6.6\phantom{)}$ & $\phantom{(}\phantom{\mathopen{}-}26.7\pm19.7\phantom{)}$ & $\phantom{(}\phantom{0}\mathopen{}-2.6\pm1.9\phantom{)}$ & $\phantom{(}0.4007\pm0.0018\phantom{)}$ & $\phantom{(}9.065\pm0.017\phantom{)}$ & $\phantom{(}50.135\pm0.072\phantom{)}$ & $\phantom{(}0.075\pm0.016\phantom{)}$\\
$\phantom{(}2454557.719757\phantom{)}$ & $\phantom{(}\phantom{\mathopen{}-}159.6\pm3.7\phantom{)}$ & $\phantom{(}\mathopen{}-11498.1\pm\phantom{0}8.3\phantom{)}$ & $\phantom{(}\mathopen{}-10.5\pm29.2\phantom{)}$ & $\phantom{(}\phantom{0}\mathopen{}-0.3\pm2.3\phantom{)}$ & $\phantom{(}0.4016\pm0.0022\phantom{)}$ & $\phantom{(}9.074\pm0.020\phantom{)}$ & $\phantom{(}50.041\pm0.080\phantom{)}$ & $\phantom{(}0.074\pm0.020\phantom{)}$\\
$\phantom{(}2454567.682293\phantom{)}$ & $\phantom{(}\phantom{\mathopen{}-}182.6\pm3.3\phantom{)}$ & $\phantom{(}\mathopen{}-11469.1\pm\phantom{0}7.5\phantom{)}$ & $\phantom{(}\phantom{\mathopen{}-}18.3\pm25.2\phantom{)}$ & $\phantom{(}\phantom{0}\mathopen{}-6.4\pm2.7\phantom{)}$ & $\phantom{(}0.4010\pm0.0020\phantom{)}$ & $\phantom{(}9.098\pm0.019\phantom{)}$ & $\phantom{(}50.259\pm0.079\phantom{)}$ & $\phantom{(}0.070\pm0.018\phantom{)}$\\
$\phantom{(}2454616.605073\phantom{)}$ & $\phantom{(}\phantom{\mathopen{}-}227.1\pm2.4\phantom{)}$ & $\phantom{(}\mathopen{}-11432.9\pm\phantom{0}7.0\phantom{)}$ & $\phantom{(}\phantom{\mathopen{}-}27.1\pm17.8\phantom{)}$ & $\phantom{(}\mathopen{}-16.4\pm2.1\phantom{)}$ & $\phantom{(}0.4084\pm0.0019\phantom{)}$ & $\phantom{(}9.042\pm0.021\phantom{)}$ & $\phantom{(}50.575\pm0.087\phantom{)}$ & $\phantom{(}0.075\pm0.017\phantom{)}$\\
$\phantom{(}2454620.562075\phantom{)}$ & $\phantom{(}\phantom{\mathopen{}-}232.3\pm3.3\phantom{)}$ & $\phantom{(}\mathopen{}-11422.1\pm10.5\phantom{)}$ & $\phantom{(}\phantom{\mathopen{}-}21.1\pm25.2\phantom{)}$ & $\phantom{(}\phantom{0}\phantom{\mathopen{}-}4.3\pm3.5\phantom{)}$ & $\phantom{(}0.4089\pm0.0029\phantom{)}$ & $\phantom{(}9.101\pm0.016\phantom{)}$ & $\phantom{(}49.615\pm0.066\phantom{)}$ & $\phantom{(}0.041\pm0.025\phantom{)}$\\
$\phantom{(}2454643.591867\phantom{)}$ & $\phantom{(}\phantom{\mathopen{}-}265.9\pm3.0\phantom{)}$ & $\phantom{(}\mathopen{}-11386.3\pm\phantom{0}8.7\phantom{)}$ & $\phantom{(}\phantom{0}\phantom{\mathopen{}-}5.3\pm23.9\phantom{)}$ & $\phantom{(}\phantom{0}\mathopen{}-7.3\pm2.8\phantom{)}$ & $\phantom{(}0.4205\pm0.0024\phantom{)}$ & $\phantom{(}9.101\pm0.016\phantom{)}$ & $\phantom{(}50.188\pm0.064\phantom{)}$ & $\phantom{(}0.053\pm0.021\phantom{)}$\\
$\phantom{(}2454878.816129\phantom{)}$ & $\phantom{(}\phantom{0}\mathopen{}-40.1\pm2.4\phantom{)}$ & $\phantom{(}\mathopen{}-11689.7\pm\phantom{0}5.7\phantom{)}$ & $\phantom{(}\phantom{0}\mathopen{}-4.9\pm17.9\phantom{)}$ & $\phantom{(}\phantom{0}\phantom{\mathopen{}-}8.7\pm2.0\phantom{)}$ & $\phantom{(}0.3961\pm0.0015\phantom{)}$ & $\phantom{(}9.188\pm0.015\phantom{)}$ & $\phantom{(}49.303\pm0.059\phantom{)}$ & $\phantom{(}0.050\pm0.014\phantom{)}$\\
$\phantom{(}2454884.801300\phantom{)}$ & $\phantom{(}\phantom{0}\mathopen{}-50.1\pm2.6\phantom{)}$ & $\phantom{(}\mathopen{}-11707.4\pm\phantom{0}7.1\phantom{)}$ & $\phantom{(}\phantom{0}\mathopen{}-9.2\pm20.3\phantom{)}$ & $\phantom{(}\phantom{\mathopen{}-}10.1\pm2.4\phantom{)}$ & $\phantom{(}0.4053\pm0.0019\phantom{)}$ & $\phantom{(}9.185\pm0.018\phantom{)}$ & $\phantom{(}49.328\pm0.070\phantom{)}$ & $\phantom{(}0.065\pm0.017\phantom{)}$\\
$\phantom{(}2454915.766283\phantom{)}$ & $\phantom{(}\mathopen{}-152.5\pm2.6\phantom{)}$ & $\phantom{(}\mathopen{}-11806.1\pm\phantom{0}6.7\phantom{)}$ & $\phantom{(}\mathopen{}-19.5\pm19.6\phantom{)}$ & $\phantom{(}\phantom{\mathopen{}-}11.1\pm2.1\phantom{)}$ & $\phantom{(}0.4050\pm0.0018\phantom{)}$ & $\phantom{(}9.196\pm0.016\phantom{)}$ & $\phantom{(}49.445\pm0.065\phantom{)}$ & $\phantom{(}0.061\pm0.016\phantom{)}$\\
$\phantom{(}2454937.689290\phantom{)}$ & $\phantom{(}\mathopen{}-209.4\pm2.3\phantom{)}$ & $\phantom{(}\mathopen{}-11871.2\pm\phantom{0}5.7\phantom{)}$ & $\phantom{(}\mathopen{}-21.6\pm17.3\phantom{)}$ & $\phantom{(}\phantom{0}\phantom{\mathopen{}-}5.0\pm1.7\phantom{)}$ & $\phantom{(}0.4005\pm0.0015\phantom{)}$ & $\phantom{(}9.135\pm0.019\phantom{)}$ & $\phantom{(}49.632\pm0.075\phantom{)}$ & $\phantom{(}0.071\pm0.014\phantom{)}$\\
$\phantom{(}2454939.703414\phantom{)}$ & $\phantom{(}\mathopen{}-235.8\pm3.5\phantom{)}$ & $\phantom{(}\mathopen{}-11889.2\pm\phantom{0}8.0\phantom{)}$ & $\phantom{(}\phantom{0}\mathopen{}-0.1\pm27.5\phantom{)}$ & $\phantom{(}\phantom{0}\phantom{\mathopen{}-}7.2\pm2.3\phantom{)}$ & $\phantom{(}0.3985\pm0.0021\phantom{)}$ & $\phantom{(}9.123\pm0.022\phantom{)}$ & $\phantom{(}49.375\pm0.088\phantom{)}$ & $\phantom{(}0.087\pm0.019\phantom{)}$\\
$\phantom{(}2454951.726105\phantom{)}$ & $\phantom{(}\mathopen{}-242.7\pm3.0\phantom{)}$ & $\phantom{(}\mathopen{}-11893.8\pm\phantom{0}7.3\phantom{)}$ & $\phantom{(}\phantom{0}\mathopen{}-9.8\pm23.2\phantom{)}$ & $\phantom{(}\phantom{0}\phantom{\mathopen{}-}5.7\pm2.2\phantom{)}$ & $\phantom{(}0.4081\pm0.0020\phantom{)}$ & $\phantom{(}9.107\pm0.018\phantom{)}$ & $\phantom{(}49.507\pm0.071\phantom{)}$ & $\phantom{(}0.071\pm0.017\phantom{)}$\\
$\phantom{(}2454993.574110\phantom{)}$ & $\phantom{(}\mathopen{}-286.1\pm2.6\phantom{)}$ & $\phantom{(}\mathopen{}-11942.3\pm\phantom{0}6.1\phantom{)}$ & $\phantom{(}\mathopen{}-53.2\pm18.9\phantom{)}$ & $\phantom{(}\phantom{0}\mathopen{}-4.4\pm1.6\phantom{)}$ & $\phantom{(}0.4007\pm0.0016\phantom{)}$ & $\phantom{(}9.074\pm0.017\phantom{)}$ & $\phantom{(}50.021\pm0.068\phantom{)}$ & $\phantom{(}0.046\pm0.015\phantom{)}$\\
$\phantom{(}2455036.504252\phantom{)}$ & $\phantom{(}\mathopen{}-250.8\pm2.6\phantom{)}$ & $\phantom{(}\mathopen{}-11902.9\pm\phantom{0}6.8\phantom{)}$ & $\phantom{(}\mathopen{}-18.8\pm20.2\phantom{)}$ & $\phantom{(}\phantom{0}\mathopen{}-7.7\pm2.2\phantom{)}$ & $\phantom{(}0.3867\pm0.0018\phantom{)}$ & $\phantom{(}9.077\pm0.018\phantom{)}$ & $\phantom{(}50.097\pm0.075\phantom{)}$ & $\phantom{(}0.068\pm0.016\phantom{)}$\\
\hline
$\phantom{(}2458117.826166\phantom{)}$ & $\phantom{(}\phantom{0}\phantom{0}\mathopen{}-0.0\pm1.7\phantom{)}$\\
$\phantom{(}2458156.839829\phantom{)}$ & $\phantom{(}\phantom{0}\mathopen{}-61.2\pm1.5\phantom{)}$\\
$\phantom{(}2458198.769910\phantom{)}$ & $\phantom{(}\phantom{0}\mathopen{}-89.5\pm1.4\phantom{)}$\\
$\phantom{(}2458257.676477\phantom{)}$ & $\phantom{(}\mathopen{}-246.1\pm2.0\phantom{)}$\\
$\phantom{(}2458488.854995\phantom{)}$ & $\phantom{(}\mathopen{}-236.5\pm1.7\phantom{)}$\\
$\phantom{(}2458540.848497\phantom{)}$ & $\phantom{(}\mathopen{}-103.5\pm1.3\phantom{)}$\\
$(2458849.842858)$ & $(\phantom{\mathopen{}-}384.7\pm144.7)$\\
$\phantom{(}2458852.845746\phantom{)}$ & $\phantom{(}\phantom{0}\mathopen{}-94.3\pm2.0\phantom{)}$\\
$\phantom{(}2458887.866178\phantom{)}$ & $\phantom{(}\mathopen{}-111.3\pm2.2\phantom{)}$\\
$\phantom{(}2459229.836282\phantom{)}$ & $\phantom{(}\mathopen{}-180.0\pm1.7\phantom{)}$\\
$\phantom{(}2459411.520342\phantom{)}$ & $\phantom{(}\phantom{\mathopen{}-}118.9\pm3.0\phantom{)}$\\
$\phantom{(}2459647.867594\phantom{)}$ & $\phantom{(}\mathopen{}-314.6\pm2.1\phantom{)}$\\
\hline
\end{tabular}}
\tablefoot{The spectra taken before and after the HARPS fiber change are separated by the horizontal line. Data points discarded due to outliers in the dLW (before 2015) or RV (after 2015) are given in parentheses.}
\end{table*}

\end{appendix}

\end{document}